\documentclass{iopjournal}
\usepackage{amsfonts,amsmath,units,wasysym,epsfig,graphicx,verbatim,color,subfigure,graphicx,bm,mathrsfs,lipsum,hyperref}
\usepackage[utf8]{inputenc}
\usepackage[normalem]{ulem}  % \sout{old text} for strikeout
\usepackage{xcolor}
\usepackage{textcase}
\usepackage{float}
\usepackage[numbers,sort&compress]{natbib}
\usepackage{chngcntr}

\newcommand{\modification}[1]{\textcolor{black}{#1}}

\begin{document}

\articletype{Article type} %	 e.g. Paper, Letter, Topical Review...

\title{A machine learning framework of ground-state properties of finite nuclei for accelerated Bayesian exploration : NucleiML}

\author{Anagh Venneti$^{1,*}$\orcid{0000-0002-0812-2702}, Chiranjib Mondal$^2$\orcid{0000-0002-9238-6144}, Sk Md Adil Imam$^3$\orcid{0000-0003-3308-2615}, Sarmistha Banik$^1$\orcid{0000-0003-0221-3651} and Bijay K. Agrawal$^{4,5}$\orcid{0000-0001-5032-9435}}

\affil{$^1$Department of Physics, Birla Institute of Technology and Science, Pilani, Hyderabad Campus, Jawahar Nagar, Kapra Mandal, Medchal District, Telangana 500078, India}

\affil{$^2$Institut d’Astronomie et d’Astrophysique, Universit\'e Libre de Bruxelles, CP 226, B-1050 Brussels, Belgium}

\affil{$^3$Instituto de Astrof\'isica, Departamento de F\'isica y Astronom\'ia, Universidad Andr\'es Bello, Santiago, Chile}

\affil{$^4$Saha Institute of Nuclear Physics, 1/AF Bidhannagar, Kolkata, 700064, West Bengal, India}

\affil{$^5$Homi Bhabha National Institute, Anushakti Nagar, Mumbai, 400094, Maharashtra, India}

\email{p20210060@hyderbad.bits-pilani.ac.in}


\begin{abstract}
The global behavior of the nuclear equation of state (EoS) is commonly studied using data from finite nuclei (FN), heavy-ion collisions, and astrophysical observations of neutron stars (NS). The constraints derived from FN such as binding energies and charge radii play the most crucial role in shaping the EoS up to saturation density. The computational cost associated with explicitly incorporating these constraints presents a significant challenge especially when the aim is to explore the model uncertainties rather than optimizing a single model. We address this by introducing NucleiML (NML), a machine learning framework trained on ground-state properties of a few finite nuclei generated by a relativistic mean-field model. NML allows us to integrate FN and NS properties within a Bayesian inference framework in an efficient manner. The results demonstrate reasonable accuracy and \modification{a speedup of $\sim 10^4$ times for calculation of FN properties for a single parameter set, yielding roughly $\sim 10^3 \times$ speed up in the Bayesian framework. The present study makes the case for extending the work to a larger set of nuclei, potentially enabling future studies of NS properties to incorporate the whole nuclear chart.}
\end{abstract}

\section{Introduction}
The behavior of the equation of state (EoS) for dense baryonic  matter plays a crucial role in determining the structural properties of neutron stars (NSs) and understanding the density dependence of nuclear symmetry energy\cite{Tsang2020,Patra2023}. The nuclear EoSs are generally parameterized using empirical values for the binding energy per nucleon, the incompressibility coefficient for symmetric nuclear matter, and the symmetry energy coefficient along with its slope at the saturation density ($\rho_0 \simeq 0.16$ fm$^{-3}$). Usually, the impact of nuclear matter parameters (NMPs) on NS properties is examined by varying them independently within ranges determined by calibration of different mean-field models using experimental  data on finite nuclei (FN). This calibration relies on measurements of binding energies\cite{Wang_2021}, charge radii\cite{angeli2013table}, isoscalar giant monopole resonances (ISGMR)\cite{Dutra2012} and other FN properties for different nuclei. Often, FN constraints are implicitly included in the analysis by imposing limits on symmetry energy and symmetry pressure, as well as pressure of symmetric nuclear matter \cite{tsang2023determination}.

The NS properties, such as mass, radius, and tidal deformability, have been inferred through pulsar observations \cite{fonseca2021refined,reardon2024neutron}, NICER\cite{riley2019nicer,riley2021nicer,miller2019psr,miller2021radius} and LIGO-VIRGO-KAGRA \cite{abbott2018gw170817rad} collaborations. These structural properties provide valuable insights into the internal composition of NSs and consequently on the behavior of highly asymmetric nuclear matter at high densities (2-8$\rho_0$). Heavy-ion collisions (HIC) data, incorporated through the values of nuclear symmetry energy and nuclear saturation properties\cite{Tsang2009, Morfouace2019, Estee2021,tsang2023determination} constrain the behavior of nuclear matter at densities 0.15-2$\rho_0$. The FN properties, such as binding energy \cite{Wang_2021}, charge radius\cite{angeli2013table}, and giant resonances\cite{Dutra2012}, measured in nuclear physics experiments, are also crucial in understanding the properties of dense nuclear matter.

Recently\cite{venneti2024unraveling}, the constraints on binding energies and charge radii of $^{40}$Ca and $^{208}$Pb nuclei were explicitly incorporated alongside those from HICs and astrophysical observations of NS properties, within a relativistic mean-field (RMF) model. This study highlights the crucial role of explicit FN constraints in understanding the NS properties. The posterior distributions obtained in this study of explicit constraints revealed a distinct correlation pattern between the NMPs and NS properties, which differs from those inferred through implicit FN constraints\cite{tsang2023determination}. The use of RMF model \cite{Gambhir1989,Gambhir1990,RING1997} in the analysis introduces a significant computational complexity, which poses challenge in extending the study to a diverse set of nuclei.

Machine learning (ML) methods have become indispensable in our modern data-driven world, where exact models are often elusive. ML has also found widespread applications in fundamental physics research. In astrophysics, it has been used to identify signatures of various gravitational wave signals from binary NS (BNS) mergers \cite{dax2025real} and supernovae \cite{abylkairov2025evaluating, mitra2024probing}, reconstruct gamma-ray burst (GRB) light curves \cite{Manchanda:2024eq1, Dainotti:2025xoe1}, recognize universal relations among neutron star properties \cite{papigkiotis2023universal, Papigkiotis:2025cjy1}, detect potential dark matter signatures in neutron stars \cite{thakur2024towards}, and classify NS EoSs \cite{gonccalves2023machine}. Due to the complexity of nuclear and particle physics, ML holds significant promise for both theoretical modeling and experimental studies \cite{hatfield2021data, benelli2022data}. In particular, ML techniques have gained considerable attention in global nuclear mass models \cite{dellen2024predicting, mumpower2023bayesian, wu2024machine, yuksel2024nuclear}. They have also been extensively used in deriving suitable energy density functionals (EDFs) along with nuclear models \cite{yang2024data, hizawa2023analysis, wu2024machine, munoz2024discovering}. Furthermore, ML-based statistical tests have also been applied to nuclear models \cite{bollapragada2020optimization, giuliani2024model}, as well as utilized to infer the EoS of NS matter \cite{ferreira2021unveiling, fujimoto2024uncertainty}. The potential of ML tools in nuclear physics has been extensively reviewed in recent work \cite{Jiang2024, y8kt-mgf5}. Their growing importance, especially in nuclear astrophysics, has been emphasized, along with the necessity for caution when applying ML techniques to extrapolate beyond known data \cite{goriely2023progress, patra2025inferring}. 

\modification{Recent efforts to accelerate FN calculations have explored both reduced-order\cite{10.3389/fphy.2022.1054524,PhysRevC.106.054322} and data analytical\cite{PhysRevLett.114.122501, Higdon_2015} strategies. Projection-based approaches such as reduced-basis methods\cite{10.3389/fphy.2022.1054524,PhysRevC.106.054322} have demonstrated substantial reduction in evaluation time for each parameter set. Simultaneously, data-driven surrogates\cite{mvc3-qdtc} have also been developed to approximate global nuclear observables and EDF predictions with significant improvement in computational speedup. While these advances have markedly reduced the cost of individual model evaluations, large-scale Bayesian analyses, which require millions of likelihood computations across multiple nuclei, still demand a framework that is both computationally efficient and easily retrained as datasets evolve. This motivates the adoption of artificial neural networks (ANNs), which are suited to learn high-dimensional nonlinear mappings directly from data and can provide rapid evaluations without reformulating the underlying physics.}

In this work, we introduce NucleiML (NML), a novel machine learning tool that incorporates a neural network-based framework. The model is trained using diverse data set comprising of various nuclei and their corresponding FN properties, derived from calculations using the RMF theory for different NMPs. The performance of NML is evaluated by comparing its predictions against RMF model computations for both nuclei included in the training set and those not encountered during the training. The NML is also integrated into a Bayesian inference framework, where the resulting posterior distributions exhibit remarkable agreement with those obtained from a similar analysis using the RMF model\cite{Gambhir1989,Gambhir1990,RING1997}.

The present paper is structured as follows. Section \ref{Sec1} provides a brief overview of the RMF formalism employed to construct the training data for NucleiML. Section \ref{Sec2} details the schematic, training, and evaluation of NML. In Section \ref{Sec3}, NML is applied to a Bayesian analysis and the resulting posteriors are compared with those obtained using the RMF model. Finally, Section \ref{Sec4} presents the summary and outlook.

\section{Relativistic Mean Field Model}\label{Sec1}
We use the relativistic mean-field (RMF)\cite{Dutra2014,Zhu_2023,nikvsic2014dirhb} model to calculate the properties such as binding energy and charge radius for a given nucleus $^A$X$_Z$. For a RMF model, nuclear matter is described by a Lagrangian, where nucleons interact with an exchange of the short-range attractive $\sigma$ mesons, very short range repulsive $\omega$ mesons, and $\rho$ mesons mediating isospin dependent interactions. In addition to mediating the interactions, the mesons also have self-interactions (in the case of $\sigma$ and $\omega$ mesons) and cross-interactions (between $\omega$ and $\rho$ mesons). The Lagrangian is defined as,
 \begin{equation}\label{eqn:1}
   \mathcal{L}_{NL} = \mathcal{L}_{nm} + \mathcal{L}_{\sigma} + \mathcal{L}_{\omega} + \mathcal{L}_{\rho} + \mathcal{L}_{int},
\end{equation}
where,
\begin{eqnarray}
    \mathcal{L}_{nm}&=& \!\bar{\psi}\left(i\gamma^\mu\partial_\mu - m\right)\psi +\!\!g_\sigma \sigma \bar{\psi} \psi -\!\!g_\omega \bar{\psi}\gamma^\mu\omega_\mu\psi 
    \nonumber \\
     &&-\frac{g_\rho}{2}\bar{\psi}\gamma^\mu\vec{\rho}_\mu \vec{\tau}\psi,\nonumber \\
%\begin{eqnarray}
    \mathcal{L}_\sigma&=&\frac{1}{2}\left(\partial^\mu\sigma\partial_\mu\sigma - m^2_\sigma \sigma^2\right) - \frac{A}{3} \sigma^3 - \frac{B}{4} \sigma^4 \nonumber,\\
%\end{eqnarray}
%\begin{eqnarray}
    \mathcal{L}_\omega&=&\! -\frac{1}{4} \Omega^{\mu\nu}\Omega_{\mu\nu} + \frac{1}{2}m^2_\omega\omega^\mu\omega_\mu +\frac{C}{4}\left(g_\omega^2 \omega_\mu \omega^\mu\right)^2 \nonumber,\\
%\end{eqnarray}
%\begin{eqnarray}
    \mathcal{L}_\rho&=& -\frac{1}{4}\vec{B}^{\mu\nu}\vec{B}_{\mu\nu} + \frac{1}{2}m^2_\rho \vec{\rho}_\mu \vec{\rho}^{ \mu} \nonumber, \\
%\end{eqnarray}
%and
%\begin{eqnarray}
    \mathcal{L}_{int}&=&\frac{1}{2} \Lambda_v g^2_\omega g^2_\rho \omega_\mu \omega^\mu \vec{\rho}_\mu \vec{\rho}^\mu.
    \label{lagrangian}
\end{eqnarray}
Here, $\Omega_{\mu\nu} = \partial_\nu \omega_\mu - \partial_\mu \omega_\nu$ and $\vec{B}_{\mu\nu} = \partial_\nu \vec{\rho}_\mu - \partial_\mu \vec{\rho}_\nu - g_\rho \left(\vec{\rho}_\mu \times \vec{\rho}_\nu \right)$. The masses of the nucleon, $\sigma$, $\omega$ and $\rho$ mesons are denoted by $m$, $m_\sigma = 508.1941$, $m_\omega = 782.501$, and $m_\rho=763.00$ respectively. The NMPs such as energy per particle of symmetric matter $E_0$, isoscalar incompressibility $K_0$, isoscalar skewness $Q_0$, Dirac effective mass of nucleons $m^*/m$, symmetry energy $J_0$, and symmetry slope parameter $L_0$, all evaluated at saturation density $\rho_0$, determine the coupling parameters $g_\sigma$, $g_\omega$, $A$, $B$, $C$, $g_\rho$ and $\Lambda_v$ \cite{Dutra2014,Zhu_2023}. The ranges of NMPs used in this study are listed in Table \ref{tab:rmfparameters}.

\begin{table}[h]
\caption{\label{tab:rmfparameters} Ranges of NMPs used to compute finite nuclei properties for closed-shell spherical nuclei within RMF theory.}
% Also listed are the range of characteristic length scale $b_{0f/m}$, value of which is different for different nuclei mass number $A$. The number of HO shells $N_{f/m}$ is also listed along with the mixing parameter $x_{\text{mix}}$ which governs the balance between stability and convergence speed in the iterative solution. The subscripts $f$ and $m$ denote fermion and meson fields
\centering
\begin{tabular}{p{3.5cm}p{3.5cm}p{1cm}}
\hline
Parameters&Values&Units\\
\hline
\hline
$\rho_0$&0.14 - 0.17 & fm$^{-3}$\\
$E_0$&-16.5 -  -15.5 & MeV\\
$K_0$&150 - 300 & MeV\\
$Q_0$&-1500 - 400 & MeV\\
$m^*/m$&0.5 - 0.8 & -\\
$J_0$&20 - 40 & MeV\\
$L_0$&20 - 100 & MeV\\
\hline
\end{tabular}
\end{table}

The computational algorithm for determining the FN properties, given a set of nuclear matter parameters (NMPs) and a specific spherical nucleus $^A$X$_Z$, can be outlined within the framework of the RMF theory as follows\cite{Gambhir1989,Gambhir1990,RING1997}:
\begin{itemize}
    \item The NMPs are utilized to determine the coupling parameters of the Lagrangian (Eq. \ref{eqn:1}), from which the field equations for mesons, photons, and nucleons are derived.
    \item These field equations are then solved by expanding the nucleon and meson fields in harmonic oscillator basis, yielding the basis occupation numbers, $n_i$, along with the corresponding meson and spinor field values.
    \item The obtained field values and occupation numbers, $n_i$, are subsequently used to compute the binding energy and charge radius.
\end{itemize}
% \av{In addition to the NMPs, the calculation of FN properties is influenced by computational parameters such as the characteristic length scale $b_{0f/m}$ and the truncation parameter $N_{f/m}$, where the subscripts $f$ and $m$ denote fermion and meson fields.} 

The RMF approach offers advantages such as naturally incorporating spin–orbit interactions and modeling nuclear forces via meson exchange, enabling a good description of bulk nuclear properties. Nevertheless, its applicability has limitations. At sub-saturation densities, nucleons remain largely non-relativistic, questioning the necessity of a fully relativistic treatment. The effective Lagrangians are somewhat ad hoc, some employing density-dependent couplings \cite{Gogelein:2007qa, Typel:1999yq} while others introduce arbitrary meson–meson interaction terms \cite{Todd-Rutel:2005yzo}. The link to underlying QCD dynamics also remains indirect. Finally, RMF models still lag behind state-of-the-art non-relativistic approaches in reproducing global mass tables with comparable precision \cite{Lalazissis:2009zz, Long:2006dj, Pena-Arteaga:2016clz, Grams:2023sml}.

%\CM{The RMF approach, with certain advantages like in-built spin-orbit interactions or the description of the nuclear interaction though a mediator, can describe bulk properties of many nuclei very well. However, inside nuclei at densities below saturation, the particles remain far away from being relativistic. The functional forms of the effective Lagrangian is also somewhat ad hoc, some with density dependent coupling constants \cite{Gogelein:2007qa, Typel:1999yq} and some other with arbitrary intra and inter-meson couplings \cite{Todd-Rutel:2005yzo}. The connection of the model to the underlying QCD dynamics also remains indirect. Furthermore, on a global scale, the accuracy on the mass table achieved with RMF models still remain quite far away from what has been achieved with non-relativistic models, cf Refs \cite{Lalazissis:2009zz,Long:2006dj,Pena-Arteaga:2016clz,Grams:2023sml} and the references therein.}

%\sout{\av{The RMF approach, while successful in describing bulk nuclear properties, has systematic limitations. It relies on the mean-field approximation and thus neglects many-body correlations. The functional form of the effective Lagrangian is somewhat ad hoc, with parameters fitted mainly to finite nuclei near stability, making extrapolations to extreme isospin or high-density regimes uncertain. Moreover, pairing, clustering, and collective dynamics are not treated explicitly, and the connection of the model to the underlying QCD dynamics remains indirect.}}

\section{\NoCaseChange{{NucleiML}}}\label{Sec2}
\subsection{The algorithm}\label{schematic}
\begin{figure}
\centering
\includegraphics[width=0.95\textwidth]{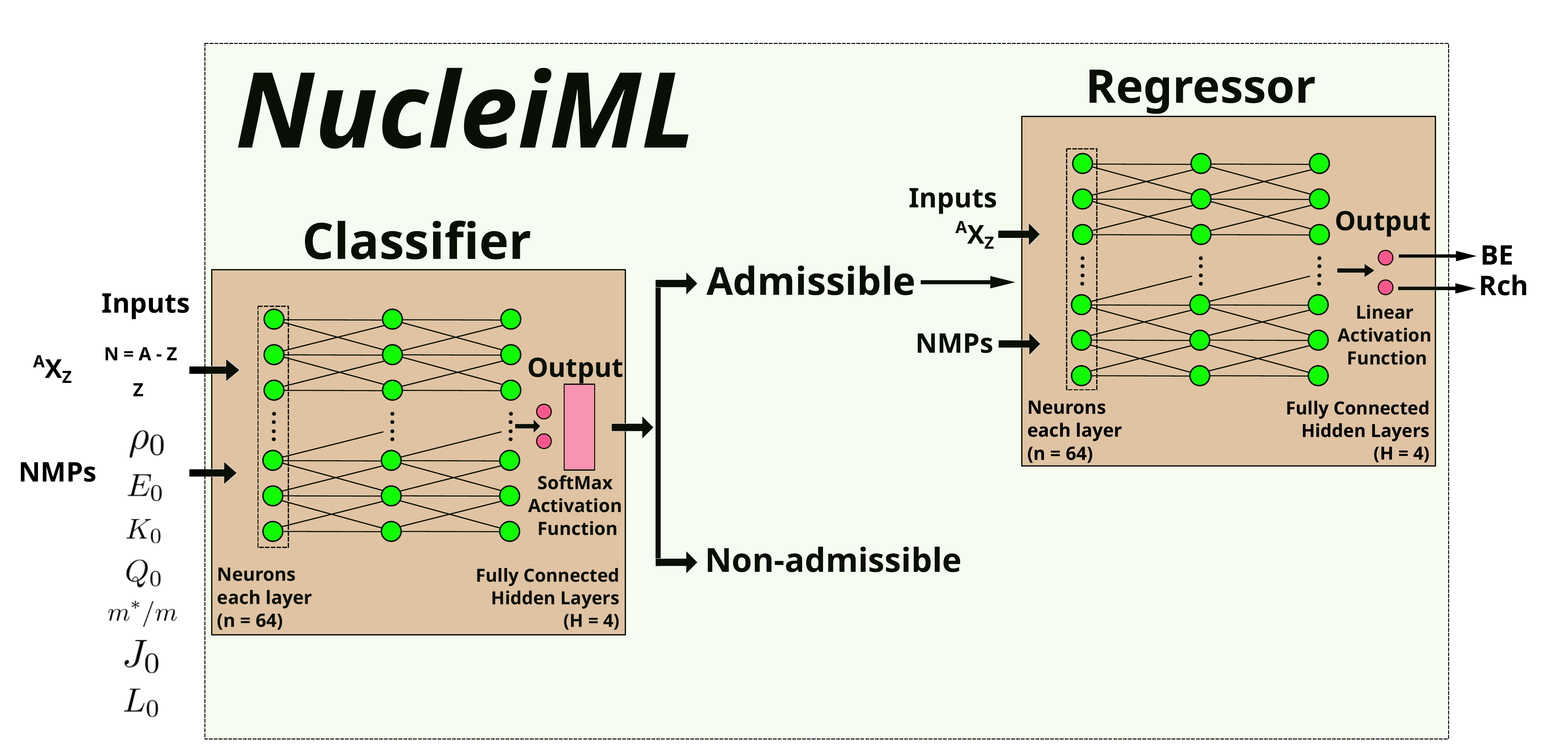}% Here is how to import EPS art
\caption{\label{flowchart} \modification{A schematic representation of NucleiML (NML), detailing the classifier and regressor. Details of the activation functions and architecture used can be found in Table \ref{tab:hyperparameters} and \ref{app_neural}.}}
\end{figure}
Statistical methods, such as Bayesian analyses, rely on random sampling of parameters and repeated model evaluations. As the number of evaluations rises, computational costs increase significantly, making it challenging to incorporate explicit FN constraints \cite{venneti2024unraveling} along with those from heavy-ion collisions\cite{tsang2023determination,Danielewicz2002, Fevre2016,Lynch2022,Tsang2009,Morfouace2019,Estee2021} and astrophysical observations\cite{fonseca2021refined,riley2019nicer,riley2021nicer,miller2019psr,miller2021radius,abbott2018gw170817rad} for a large scale sampling. Machine learning techniques, particularly neural networks, provide an effective solution by improving computational efficiency while emulating the RMF model.

\modification{NucleiML (NML) is a neural network-based framework designed to emulate the predictions of the RMF model while improving computational speed and preserving close agreement with the underlying model.} As illustrated in Fig.\ref{flowchart}, the NML algorithm follows a two-step implementation process. Firstly, an input set of NMPs and a nucleus $^A$X$_Z$ are classified into two of categories: \textit{admissible} and \textit{Non-admissible}. The classes are determined from the behavior of the inputs. The \textit{admissible} class is the input category which is determined to yield convergent results for finite nuclear properties. The \textit{non-admissible} inputs are those that do not yield a valid set of coupling parameters, convergent FN properties or FN properties that are within 20\% of the experimentally determined values.

\begin{table}
\caption{\label{tab:hyperparameters} The hyper-parameters of the neural network, estimated by a grid search, for which the loss is minimized, thereby maximizing the accuracy of the neural network.}
\centering
\begin{tabular}{lcccc}
\hline
NN Model & Epochs &  Layers & Neurons& learning rate\\
\hline
\hline
NML Classifier&800&4&64& 0.0001\\
NML regressor&500&4&64& 0.001\\
\hline
\end{tabular}

\end{table}
In the second step, a neural network regressor is employed for calculating the predicted binding energies and charge radii of $^A$X$_Z$, for those inputs categorized \textit{admissible} by the NML classifier. The neural networks are built and trained using the python packages of \textit{Tensorflow}\cite{tensorflow2015-whitepaper} and \textit{Keras}\cite{chollet2015keras}.

A crucial step in optimizing the performance of the neural network is the hyper-parameter estimation. This is done by conducting a grid search to optimize the number of layers, neurons per layer, and training epochs. For this purpose, we use the Sklearn and \modification{SciKeras} packages\cite{scikit-learn} to identify the optimal configuration of hyperparameters, that minimize losses while maximizing accuracy, by iterating over a grid of all possible hyperparameters.

Different hyperparameters are usually employed to capture the diverse patterns of the underlying model. In this study, we primarily vary three hyperparameters while keeping the others fixed. The number of training epochs is chosen to balance efficiency with accuracy, preventing both underfitting from too few epochs and overfitting from an excessively large number. The depth of the network (number of layers) and the number of neurons per layer are selected to capture the necessary model complexity and data variability, while also controlling computational cost. The learning rate controls the speed and stability during the optimization of weights in training. Table \ref{tab:hyperparameters} summarizes the hyper-parameters determined by the grid search for the NML classifier and regressor. Further discussion on the significance of other hyperparameters, along with additional hardware specifications are provided in \ref{app_neural}.

\subsection{The Classifier}
\subsubsection{Training}
We construct a large data set by randomly sampling the seven NMPs, which are then utilized to determine the seven coupling constants of the Lagrangian in Eq.\ref{eqn:1}. These coupling parameters serve as inputs for computing the ground state  properties of five spherically symmetric closed shell nuclei: $^{16}$O$_{8}$, $^{40}$Ca$_{20}$, $^{48}$Ca$_{20}$, $^{132}$Sn$_{50}$, and $^{208}$Pb$_{82}$, within the RMF model. The data set includes NMPs, coupling parameters, corresponding binding energy (BE), charge radius ($R_{ch}$) and the class (whether \textit{admissible} or \textit{Non-admissible}) for a nucleus, $^A$X$_Z$. \modification{The NMPs, together with the nucleus identifiers $N$ and $Z$, serve as inputs to both the classifier and the regressor. The classifier maps these inputs to a class label (admissible or non-admissible), while the regressor maps them to the corresponding FN observables.}

Before being fed into the NML classifier, the inputs, NMPs and nucleus $^A$X$_Z$, undergo scaling and normalization. This step ensures that widely varying parameter values do not unduly impact the training process of the neural network. \modification{The dataset is then divided into three subsets: training, validation, and testing sets, using a 70\%–15\%–15\% split, respectively.} The classifier is trained to minimize categorical cross-entropy, learning diverse classification trends of the training data set. The categorical cross-entropy loss function\cite{mao2023cross} is defined as
\begin{equation}
     H\left(p,q\right) = -\sum_i p_i \times log(q_i),
 \end{equation}
where $p_i$ and $q_i$ are the true and predicted labels respectively, summed over the training dataset. During each training epoch, the neural network is first trained on the training set and then evaluated on the validation set, assessing its ability for generalization to unseen data.

\begin{figure}[h]
\centering
\includegraphics[width=0.5\textwidth]{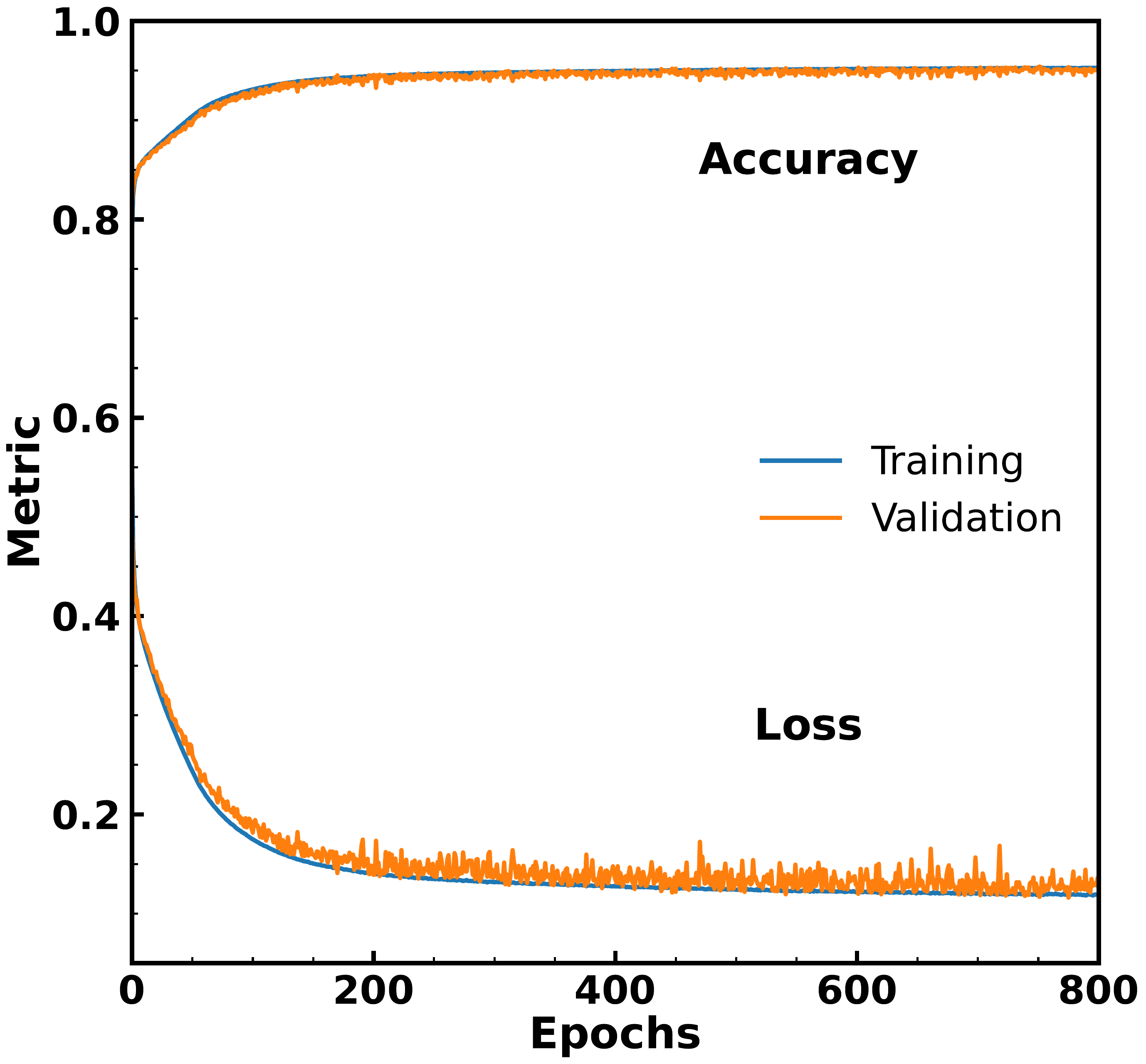}% Here is how to import EPS art
\caption{\label{fig:training_class}The training loss and accuracy of the classifier over different training epochs. The training loss progressively reduces till it is stabilized approximately around $0.2$. The accuracy also gets progressively better, stabilizing at $\sim 0.95$}
\end{figure}

As illustrated in Fig. \ref{fig:training_class}, the training loss of the NML classifier starts at a high value and gradually decreases as training progresses, eventually stabilizing around a loss of approximately $\sim 0.2$. Alongside the training loss, the validation loss is also shown, exhibiting minor fluctuations but following a similar downward trend. This behavior suggests that the NML classifier effectively generalizes to unseen data. Moreover, the absence of a significant divergence between training and validation losses indicates that the model is not over-fitting to the training data. The accuracy trends for both the training and validation sets further reinforce this observation. Over successive epochs, the accuracy increases steadily and stabilizes around 95\%, demonstrating the effectiveness of the classifier in distinguishing between the categories.

\modification{The classifier acts as a computationally efficient surrogate for physical admissibility, ensuring that Bayesian sampling remains confined to numerically stable and phenomenologically meaningful regions of parameter space. A detailed discussion of the motivation, representative examples of non-admissible parameter sets, sensitivity to threshold choices, and quantitative impact on posterior inference are presented in \ref{class}.}

\subsubsection{Performance}
We evaluated the performance of the NML classifier using metrics such as accuracy, precision, recall, and the F1 score. The overall accuracy of the classifier is 95\%, demonstrating its effectiveness in categorizing inputs, previously also observed during neural network training (Fig. \ref{fig:training_class}). Additional metrics such as precision and recall provide deeper insight into the performance of classifier across different classes. Precision measures the proportion of correctly predicted instances for a given class out of all instances predicted to belong to that class. Recall, on the other hand, quantifies the proportion of correctly classified instances among all actual instances of a particular class. The F1 score, which is the harmonic mean of precision and recall, is particularly useful for imbalanced datasets, as it provides a more balanced assessment of classification performance. Table \ref{tab:class_report} summarizes these metrics. The NML classifier performs exceptionally well in identifying the input into the two classes as indicated by reasonable total accuracy of 95\%. The strong scores for precision and recall for the classes also highlight the strength of the classifier in capturing the class wise diversity in the data.

\begin{table}
\caption{\label{tab:class_report}Summary of the performance metrics of the classifier}
\centering
\begin{tabular}{lccc}
\hline
Class&Precision&Recall&F1 Score\\
\hline
\hline
Admissible&0.93&0.97&0.95 \\
Non admissible &0.97&0.94&0.95 \\
\hline
Total Accuracy&&&0.95 \\
\hline
\end{tabular}
\end{table}

\begin{figure}[h]
\centering
\includegraphics[width=0.49\textwidth]{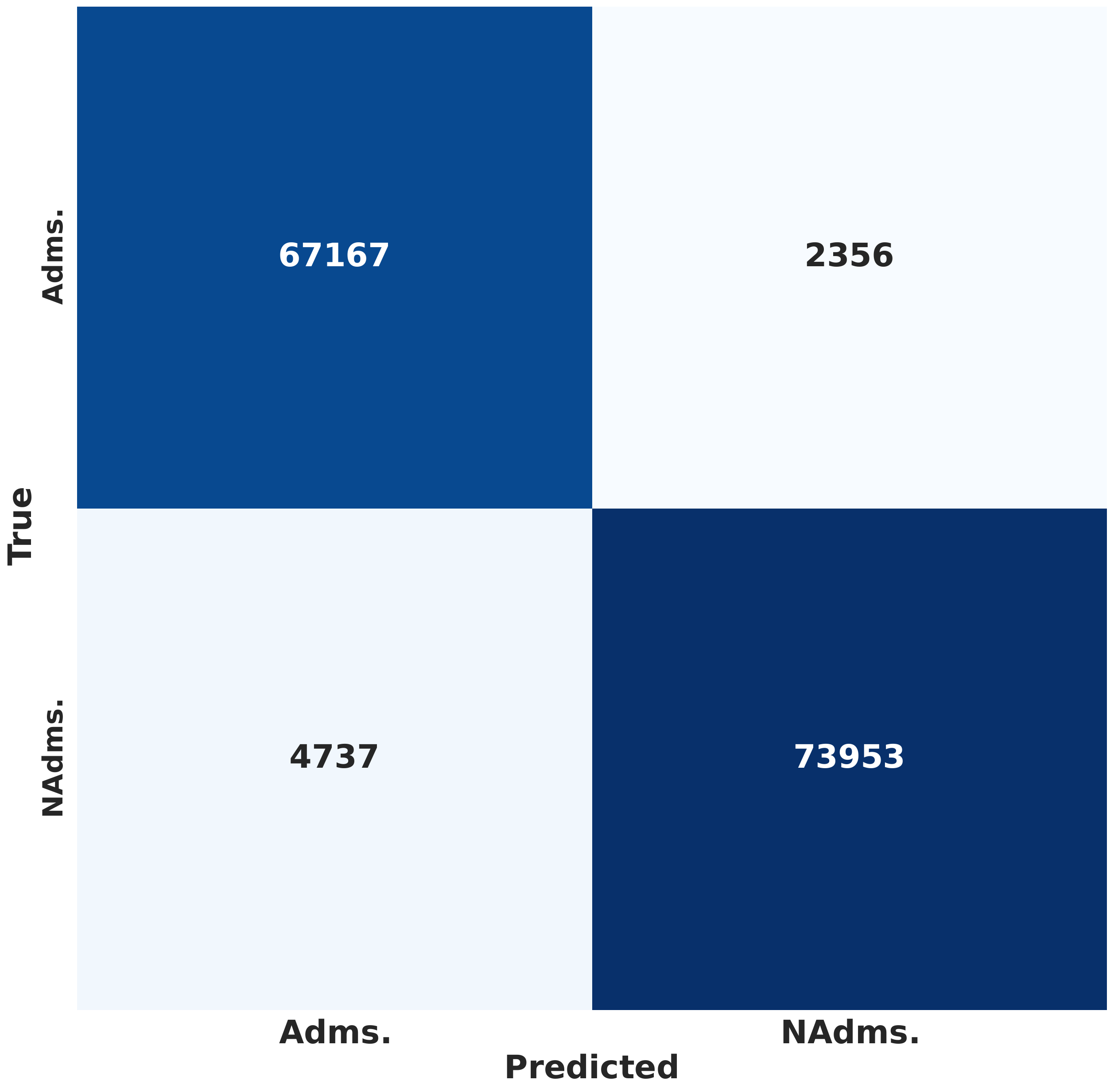}% Here is how to import EPS art
\caption{\label{fig:confusion}The true and predicted labels are shown here in a confusion matrix. The diagonal elements display the correctly classified instances, while the off diagonal elements indicate the misclassifications. Larger values, as indicated by darker shade of blue, along the diagonal suggest a strong performance of the classifier, providing insight into the class-wise accuracy. Adms. and NAdms. indicate the classes \textit{Admissible} and \textit{Non-admissible}.}
\end{figure}

Fig.\ref{fig:confusion} presents the confusion matrix, illustrating the true and predicted labels. The diagonal elements of the matrix represent correctly classified instances, with larger values indicating better classification performance. Misclassifications are reflected in off-diagonal elements. The size ($\sim 150,000$) of the test data set can be obtained by adding all the instances that appear in the confusion matrix. As previously detailed in Table \ref{tab:class_report}, the NML classifier exhibits a strong performance across the two classes, as also evidenced by the high values in the diagonal elements. The classifier is particularly valuable in controlling the acceptance rate, while sampling a Bayesian posterior distribution,to minimize the risk of sampling outliers in data. 

\begin{figure}[h]
\centering
\includegraphics[width=0.45\textwidth]{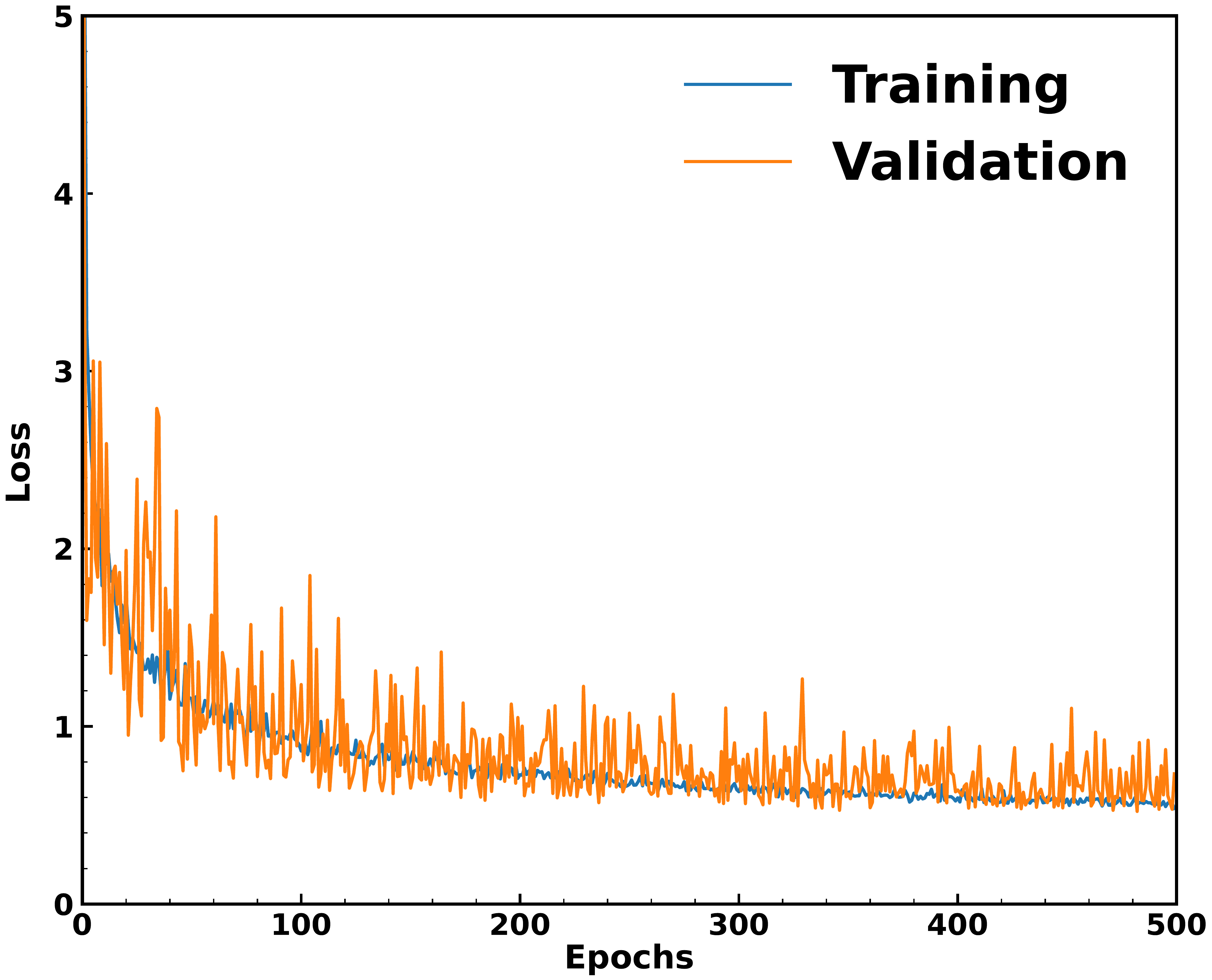}% Here is how to import EPS art
\caption{\label{fig:loss}The training of regressor model for binding energy and charge radius is shown here. The mean absolute percentage error (MAPE) loss reduces as the training of the neural network proceeds for a longer epochs. Training and validation losses show a similar trend with progressive improvement, with training loss stabilized at $\sim 0.5\%$}
\end{figure}

\subsection{The Regressor}
\subsubsection{Training}
The second component of the NML framework is the regressor. The calculation of FN properties, such as binding energy and charge radius for a given set of NMPs and a nucleus $^A$X$_Z$, is formulated as a regression problem. As with the classifier, the inputs for the regressor are scaled and normalized. \modification{The outputs, however, are left unscaled since scaling does not significantly influence the predicitve performance of the regressor. For the regressor, we employed a 70\%–21\%–9\% split of the dataset for training, validation and testing, respectively.} The neural network is trained to capture the underlying trends in the training dataset by minimizing a loss function. Specifically, we employ the mean absolute percentage error (MAPE)\cite{de2016mean} as the loss function, defined as:
\begin{equation}
    MAPE = 100\frac{1}{n}\sum^{n}_{t=1}\left|\frac{A_t-P_t}{A_t}\right|,
\end{equation}
where $A_t$ and $P_t$ are the actual and predicted value and $n$ is the total number of training data points.

Fig. \ref{fig:loss} shows the training loss decreasing steadily over epochs for both binding energy and charge radius, converging to about $\sim 0.6\%$.  \modification{The initial fluctuations in validation loss are due to randomly initialized weights during the early stages of training. The stochastic mini-batch training and the learning rate dynamics in a non-convex loss space contribute to the continued variability. Nevertheless, the overall trend demonstrates a steady reduction in loss, followed by stabilization as the number of epochs increases.The regressor achieves a global R$^{2}$ score of 0.998, computed across the complete test dataset aggregating predictions for all nuclei and both observables. This indicates that NML reproduces the overall variance structure of the aggregated dataset faithfully.}

\begin{figure}[h]
\centering
\includegraphics[width=0.45\textwidth]{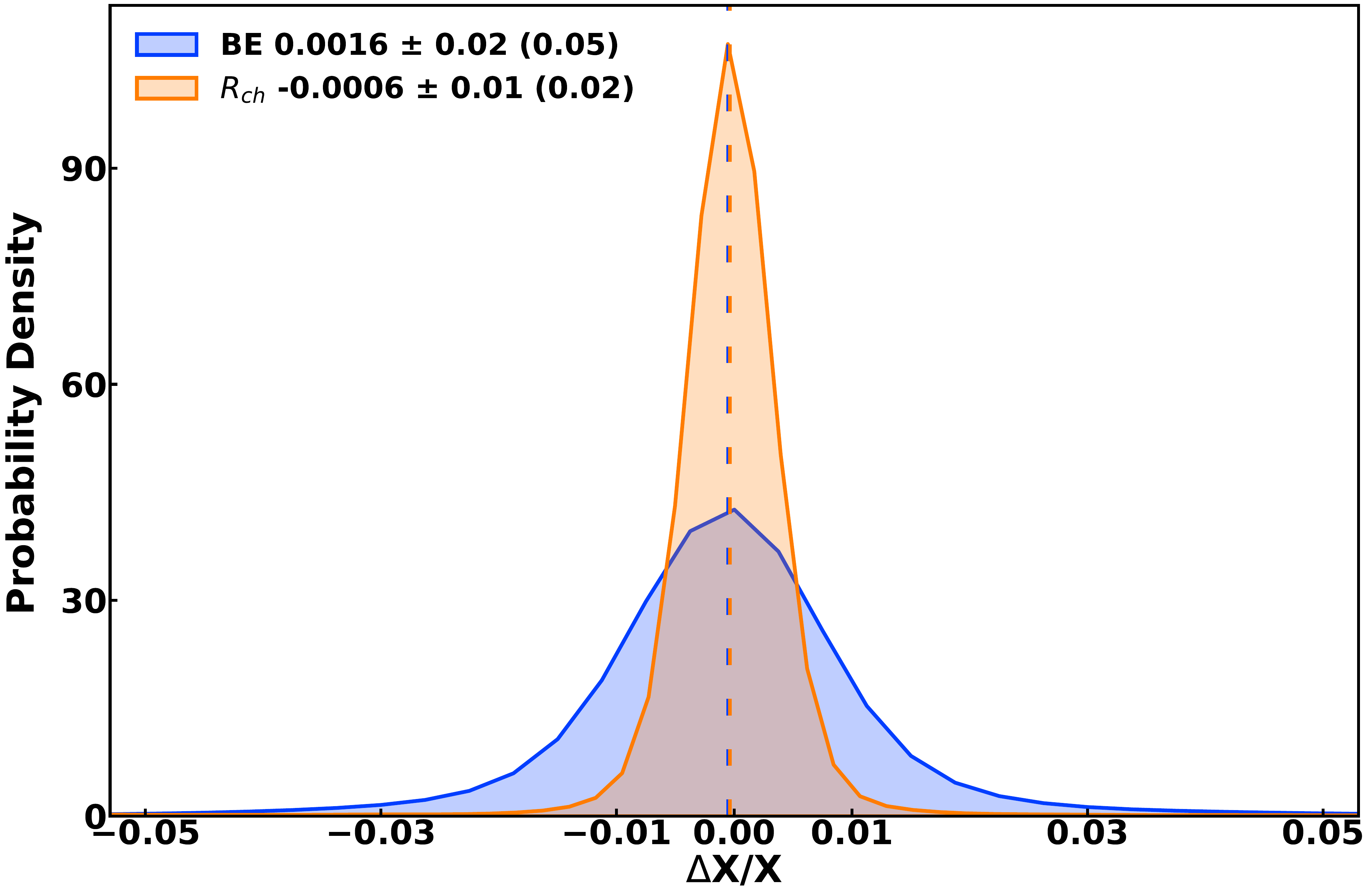}% Here is how to import EPS art
\caption{\label{fig:error_be}The percentage deviation ($\Delta X/X$) of the set of nuclei used in training, evaluated for the test dataset of NMPs, for binding energy (BE) and charge radius ($R_{ch}$). The mean and standard deviation of the distribution are shown in the legend, with 95\% CI values reported in parentheses.}
\end{figure}
\subsubsection{Performance}

In Fig. \ref{fig:error_be}, we present the probability density distribution of the deviations between the predicted and true values of binding energy and charge radius of FN, with their respective mean and standard deviation of the distributions reported in the legend. The deviation, $\Delta X/X$ is defined as,
\begin{equation}\label{dev}
    \frac{\Delta X}{X} = \frac{X_{true} - X_{pred}}{X_{true}}
\end{equation}
where X represents either the binding energy or the charge radius. The deviation for FN properties exhibits a peak around zero, with reasonably narrow distributions, highlighting the reasonable accuracy of the NML regressor. The distributions appear to be Gaussian in nature with the mean ($\mu$) centered around zero, with the spread of the distribution indicated by the standard deviation ($\sigma$) of the distribution. The binding energy exhibits a broader distribution ($\sigma \sim 0.05$) as compared to the narrower distribution ($\sigma \sim 0.02$) of charge radius, which is also reflected in the $\sigma$ and 2$\sigma$ intervals.

\begin{figure}[htbp]
    \centering
    \begin{minipage}{0.5\textwidth}
        \centering
        \includegraphics[width=\textwidth]{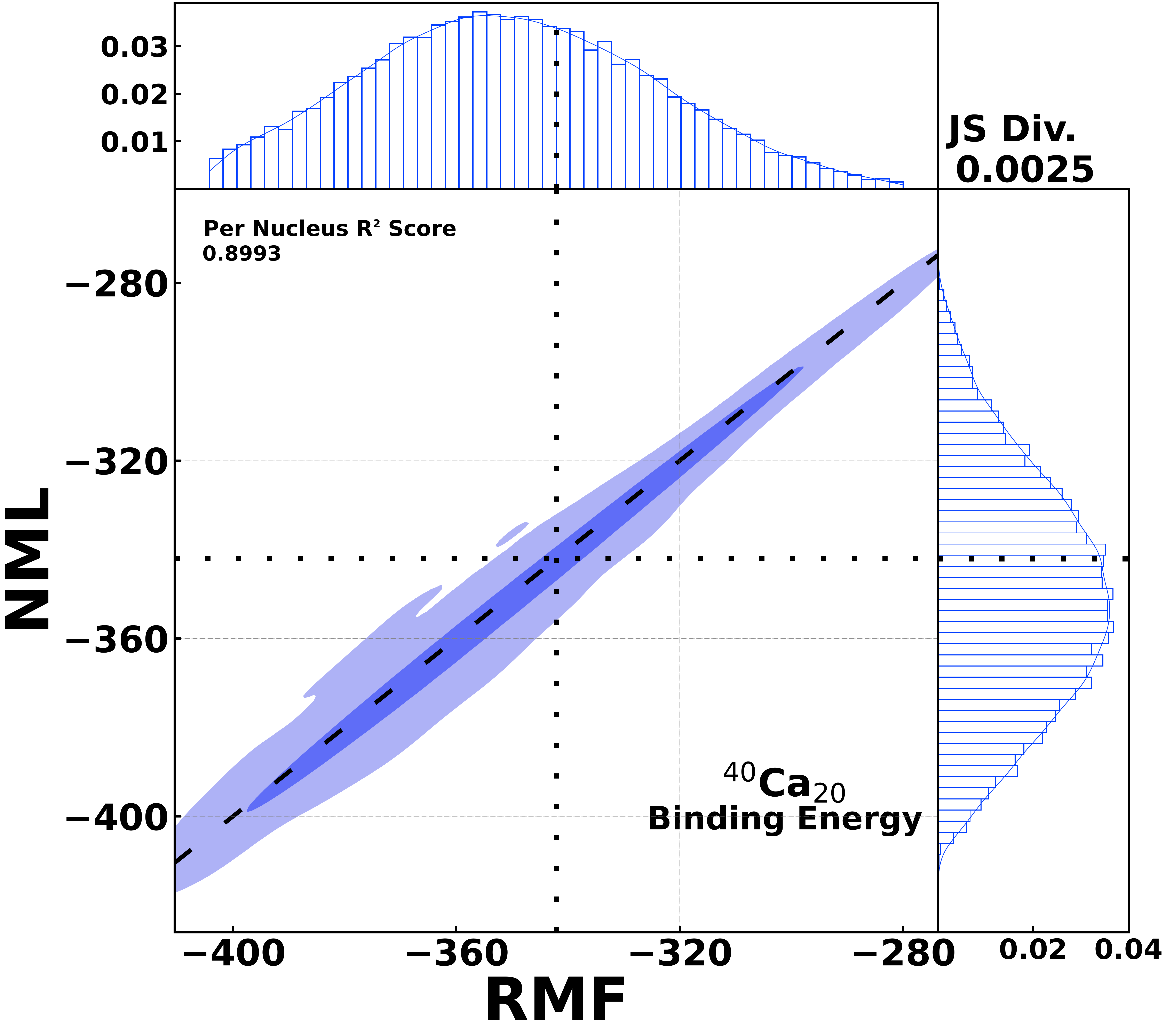}
        % \caption{Caption for Plot 1}
    \end{minipage}
    \hfill
    \begin{minipage}{0.48\textwidth}
        \centering
        \includegraphics[width=\textwidth]{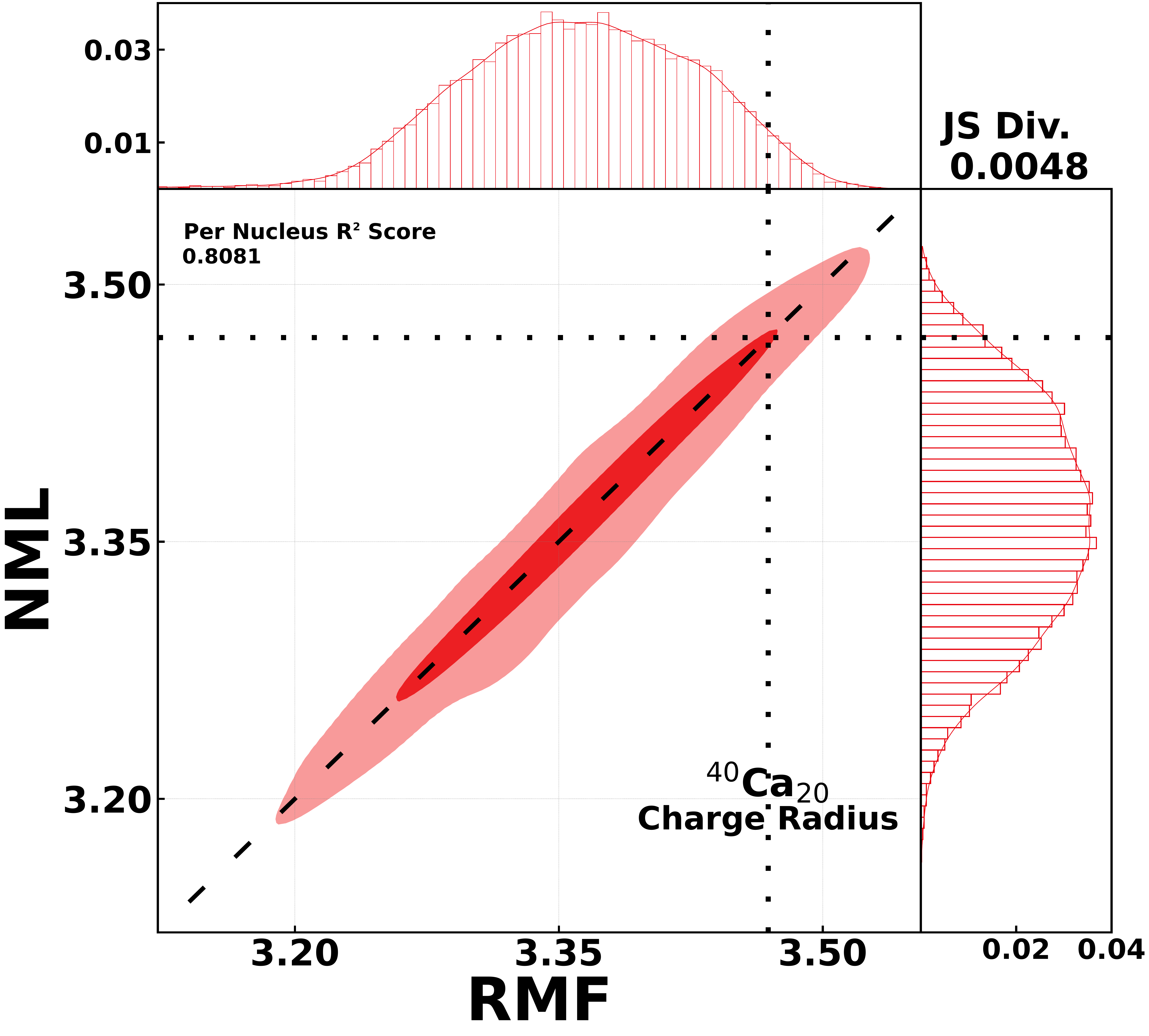}
        % \caption{Caption for Plot 2}
    \end{minipage}
    
    \vspace{1.5cm}  % Adjust vertical spacing
    
    \begin{minipage}{0.5\textwidth}
        \centering
        \includegraphics[width=\textwidth]{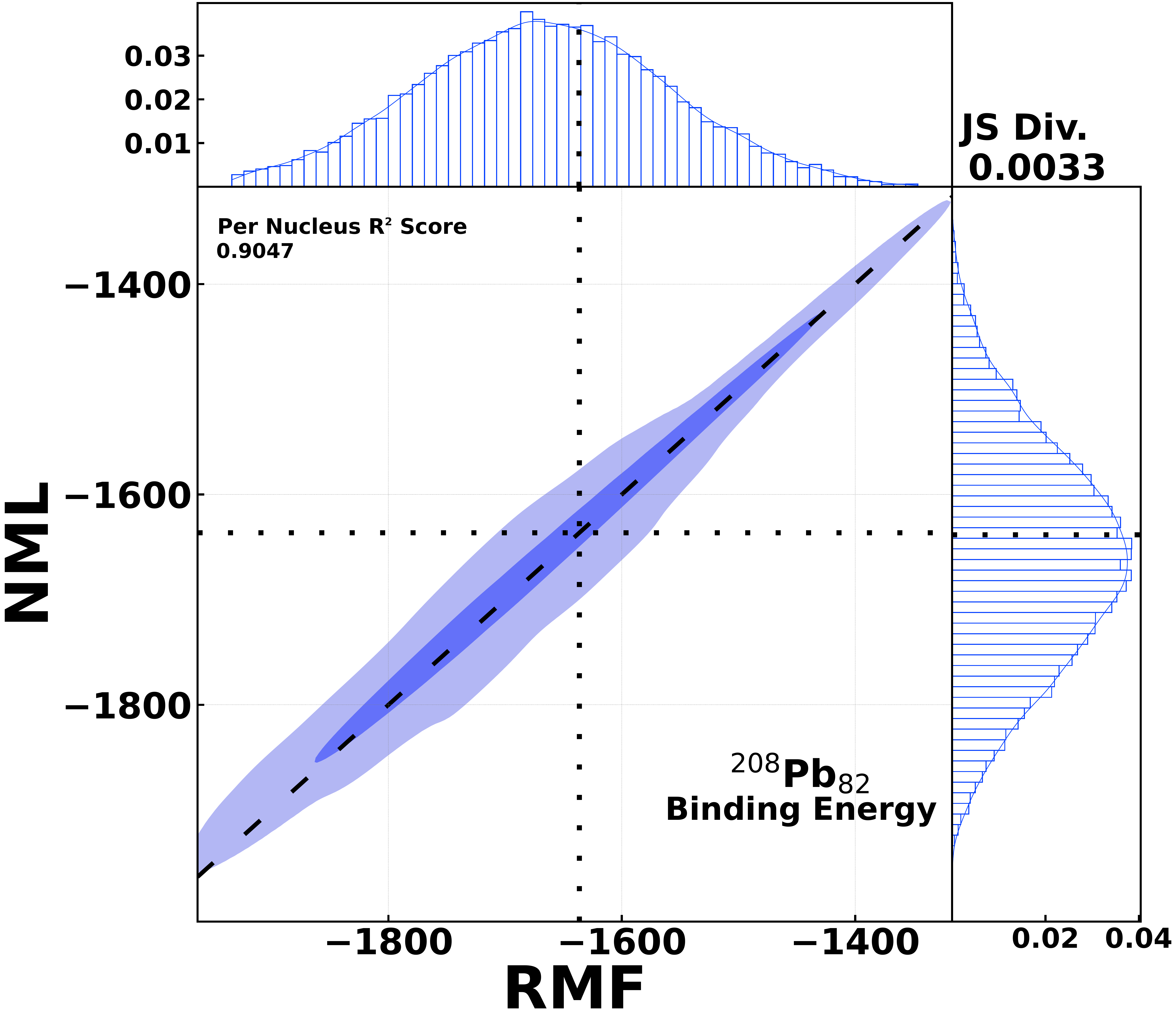}
        % \caption{Caption for Plot 3}
    \end{minipage}
    \hfill
    \begin{minipage}{0.48\textwidth}
        \centering
        \includegraphics[width=\textwidth]{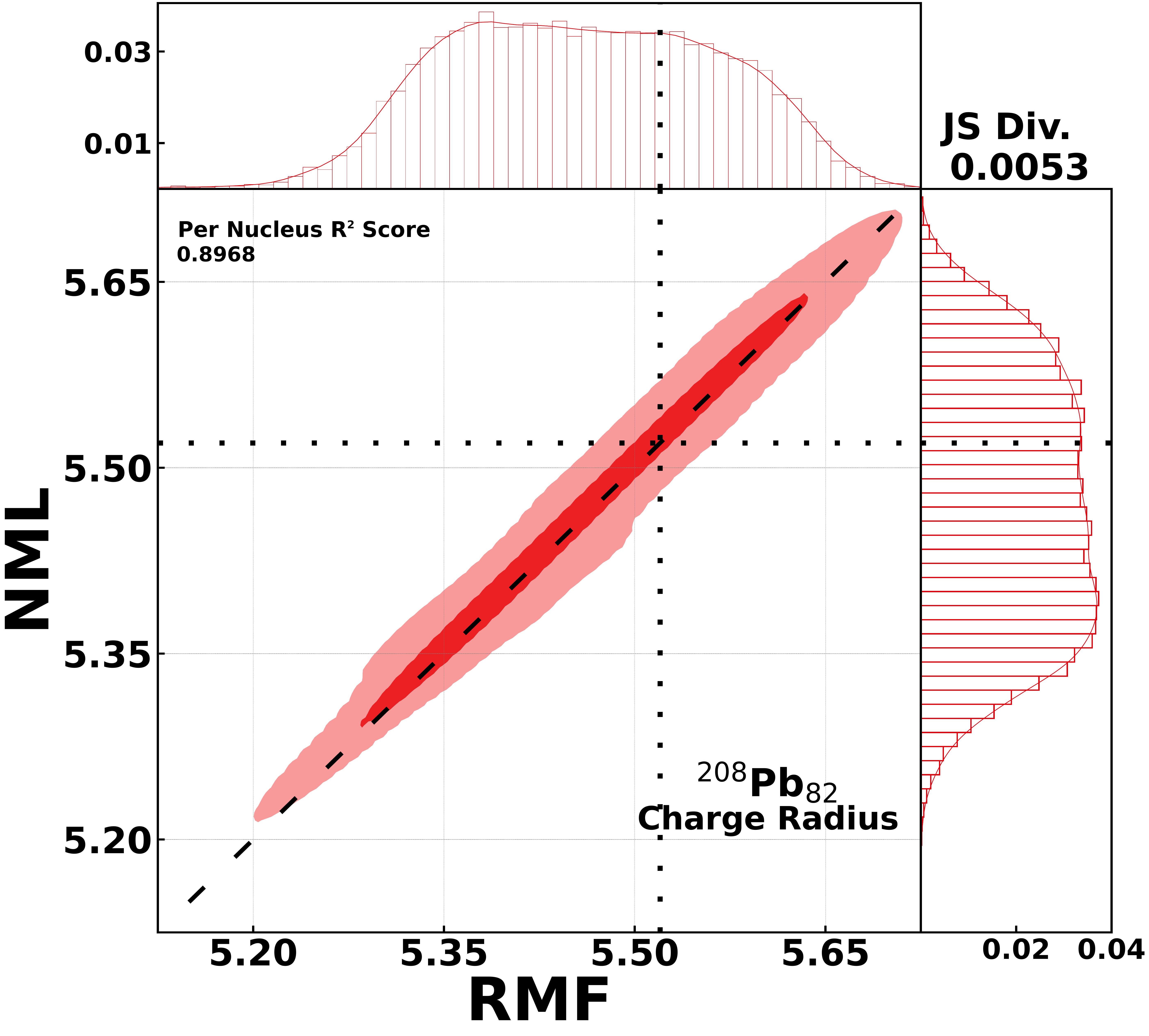}
        % \caption{Caption for Plot 4}
    \end{minipage}

    \caption{\modification{The KDE distributions of the predictions for binding energy and charge radius of $^{40}$Ca$_{20}$ and $^{208}$Pb$_{82}$ from RMF model and NucleiML. The experimental values are shown as vertical and horizontal black dotted lines. The ideal regression line is shown as a black dashed line. Additionally, the marginalized distributions of both models as well the JS divergence values of the two are displayed. The per-nucleus $\text{R}^2$ score for each of the observable is also indicated.}}
    \label{fig:all_plots}
\end{figure}
% \newpage
We analyze the performance of the NML regressors for specific nuclei, such as $^{40}$Ca$_{20}$ and $^{208}$Pb$_{82}$, in Fig.\ref{fig:all_plots}.\modification{ We present joint distributions (blue for binding energy and red for charge radius) that illustrates the distribution of FN properties computed using the RMF model alongside the corresponding predictions from the NML. The experimentally determined values are represented by vertical and horizontal black dotted lines, while the dashed black line indicates the ideal regression trend. The per-nucleus $\text{R}^2$ score for each observable is also reported.} Fig.\ref{fig:all_plots} also shows marginalized 1D posterior distributions of the FN properties obtained from RMF and NML models, along with the Jenson-Shanon (JS) divergence\cite{lin1991divergence}, which is defined as,
\begin{equation}
JS(P \parallel Q) = \tfrac{1}{2} \, KL\!\left(P \parallel M\right) 
+ \tfrac{1}{2} \, KL\!\left(Q \parallel M\right),
\end{equation}
where
\begin{equation}
M = \tfrac{1}{2}(P + Q),
\end{equation}
P, Q being the distributions of RMF and NML predictions respectively. $KL(A||B)$ is the KL divergence\cite{Kullback1951,mackay2003information} between the two distributions, quantifying the difference between the distributions.

% It can be observed that the NML effectively predicts nuclear properties, closely following the expected trend. The model exhibits reliable accuracy, with the 95\% CI of the deviation ranging from -$3\%$ to $5\%$ of the binding energy and -$1\%$ to $1\%$ for the charge radius, relative to RMF calculations. This indicates that the model reasonably approximates RMF results, with slightly greater variability in binding energy predictions compared to charge radius estimates. 

\modification{NML effectively predicts the FN properties, closely following the ideal regression trend. The per-nucleus $\text{R}^2$ scores validate this agreement, which is the highest ($\sim0.91$) for binding energy of $^{208}$Pb$_{82}$ and the lowest ($\sim 0.81$) for the charge radius of $^{40}$Ca$_{20}$. This indicates that the model reasonably approximates RMF predictions, while exhibiting greater variability at the individual nucleus level compared to the aggregated global performance reflected by the high global $\text{R}^2$ score. The JS divergence values further demonstrate strong similarity between the distributions obtained from NML and the RMF model. This complements the pointwise regression metrics by establishing distribution-level agreement with the underlying model. A detailed nucleus-wise breakdown of the predictive performance is provided in \ref{nuc_wise_rel_err}.}

\begin{figure}[h]
\centering
\includegraphics[width=0.5\textwidth]{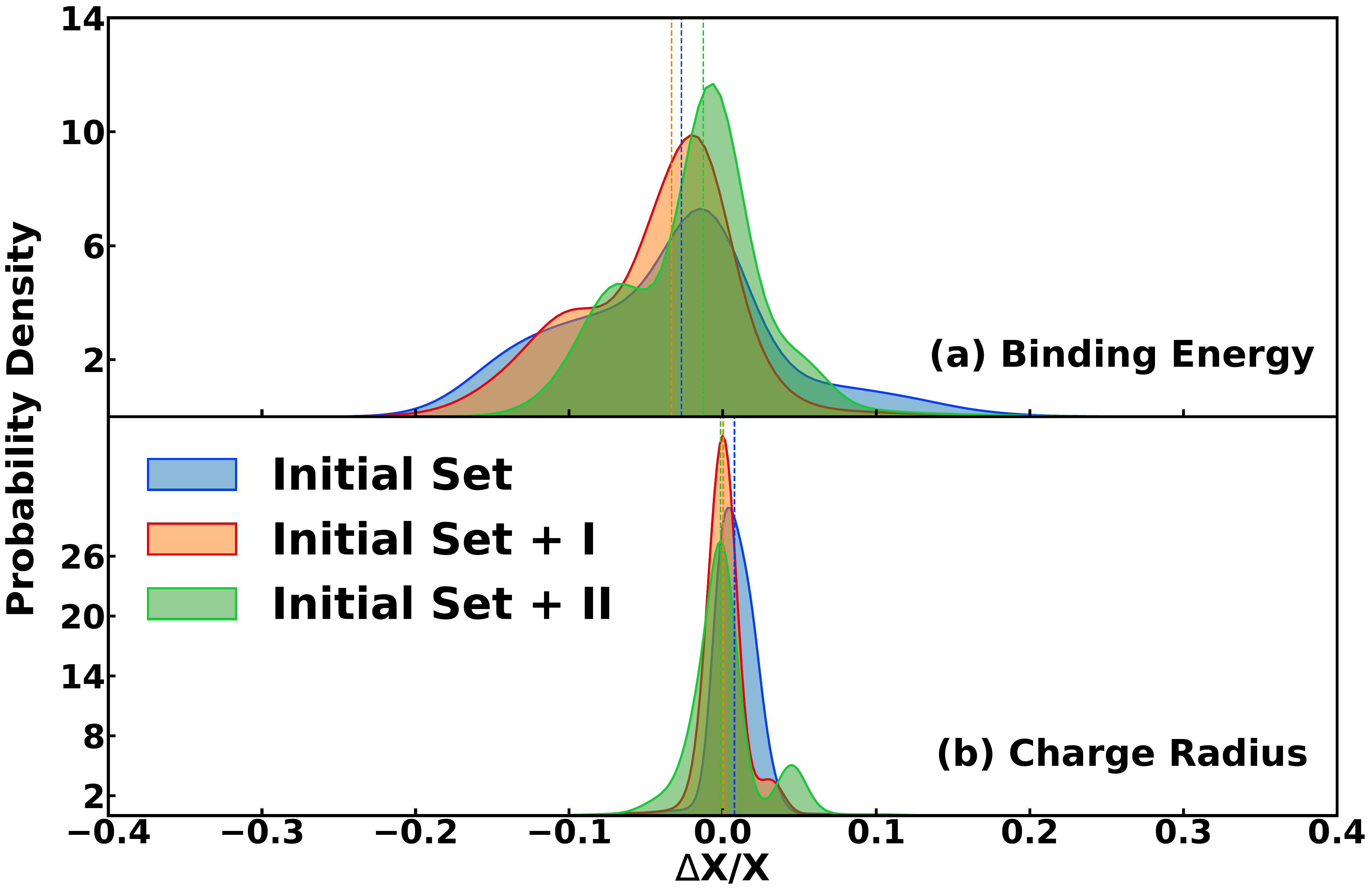}% Here is how to import EPS art
\caption{\label{fig:untrained}The percentage deviation ($\Delta X/X$) for the set of nuclei that were not used in the training of the regressor model for (a) binding energy and (b) charge radius. The nuclei used in the training are indicated by the color, with blue indicating the initial set of nuclei ($^{16}$O$_{8}$, $^{40}$Ca$_{20}$, $^{48}$Ca$_{20}$, $^{132}$Sn$_{50}$, and $^{208}$Pb$_{82}$). The orange and green indicate set I ($^{24}$O$_{8}$, $^{58}$Ca$_{20}$, and $^{78}$Ni$_{28}$) and set  II ($^{24}$O$_{8}$, $^{58}$Ca$_{20}$, $^{78}$Ni$_{28}$, $^{68}$Ni$_{28}$ along with $^{90}$Zr$_{40}$) in addition to the nuclei initially considered for training.}
\end{figure}
Fig. \ref{fig:untrained} presents the performance of the NML regressor by showing the probability distribution of the deviation $\Delta X/X$ for (a) binding energy and (b) charge radius in nuclei that were not included in the training set. For the untrained set, we consider 7 nuclei :$^{30}$Ne$_{10}$, $^{54}$Ca$_{20}$, $^{56}$Ni$_{28}$, $^{100}$Sn$_{50}$, $^{116}$Sn$_{50}$, $^{138}$Sn$_{50}$ and  $^{144}$Sm$_{62}$. As observed, the initial training set, consisting of five selected spherically symmetric closed shell nuclei: $^{16}$O$_{8}$, $^{40}$Ca$_{20}$, $^{48}$Ca$_{20}$, $^{132}$Sn$_{50}$, and $^{208}$Pb$_{82}$, exhibits a broad deviation distribution (shown in blue). This suggests difficulties in accurately predicting the properties of unseen nuclei, highlighting the need for a more diverse training dataset. To address this, we expand the training set by incorporating additional nuclei in multiple stages. First, we introduce three neutron-rich nuclei, $^{24}$O$_{8}$, $^{58}$Ca$_{20}$, and $^{78}$Ni$_{28}$, and retrain the model. The resulting distribution of the deviation (shown in orange) indicates an improvement, with the median deviation shifting closer to zero and the overall spread becoming narrower. Further improvements are observed when two more nuclei, $^{68}$Ni$_{28}$ along with $^{90}$Zr$_{40}$, are included in the training set. The median deviation aligns even closer to zero, and the distribution (shown in green) tightens significantly, demonstrating a refined accuracy. \modification{A quantitative comparison of per-nucleus $\text{R}^2$ scores for the initial and extended training sets shows that performance improves substantially. Several nuclei  show noticeable gains after inclusion of the extended set, as compated to the initial configuration (see \ref{nuc_wise_rel_err} for further details).} These results underscore the importance of training on a more diverse range of nuclei to improve the predictive performance of NML regressor. 
\subsection{Accuracy of NML}
A deviation of 5\% provides a reasonable starting point for the objective of developing a computationally efficient emulator of explicit finite nuclei constraints. Out of a  sample size of $\sim 120,000$ input parameters, only about 4000 ($\sim4\%$) exceed this deviation indicating that NML struggles only for a small fraction of inputs. Further examination of these parameters show that their distribution differs from the rest of the distribution for which NML performs reliably.

A comparison with the experimental data sheds further light into this behavior. For samples where NML deviates from RMF by more than 5\%, the RMF predictions themselves exhibit a median deviation of $\sim 8\%$ from experimental binding energies. Conversely, for samples where RMF deviates from experiment by less than 2\%, NML reproduces the RMF predictions within the deviation of about 2\%. This indicates that largest deviations of NML occur mainly for outliers, where RMF predictions are already less reliable, while it closely reproduces RMF results for well-behaved cases.

Although errors of $\sim$2\% appear large in comparison to global mass models that target keV-scale accuracy, they are acceptable for our objective of reducing the computational cost of incorporating explicit finite nuclei constraints in the study of NS EoS. The higher deviation of 5\% further ensures that the Bayesian sampling of the parameters are not distorted by sampling of outliers. Also, experimental values of charge radii are incorporated into the explicit finite nuclei constraints where the reliability of NML in reproducing the charge radii of different nuclei ($<$1\%) helps in compensating the larger uncertainty for binding energies.

\section{A sample Bayesian run}\label{Sec3}
\begin{figure}[h]
\centering
\includegraphics[width=0.9\textwidth]{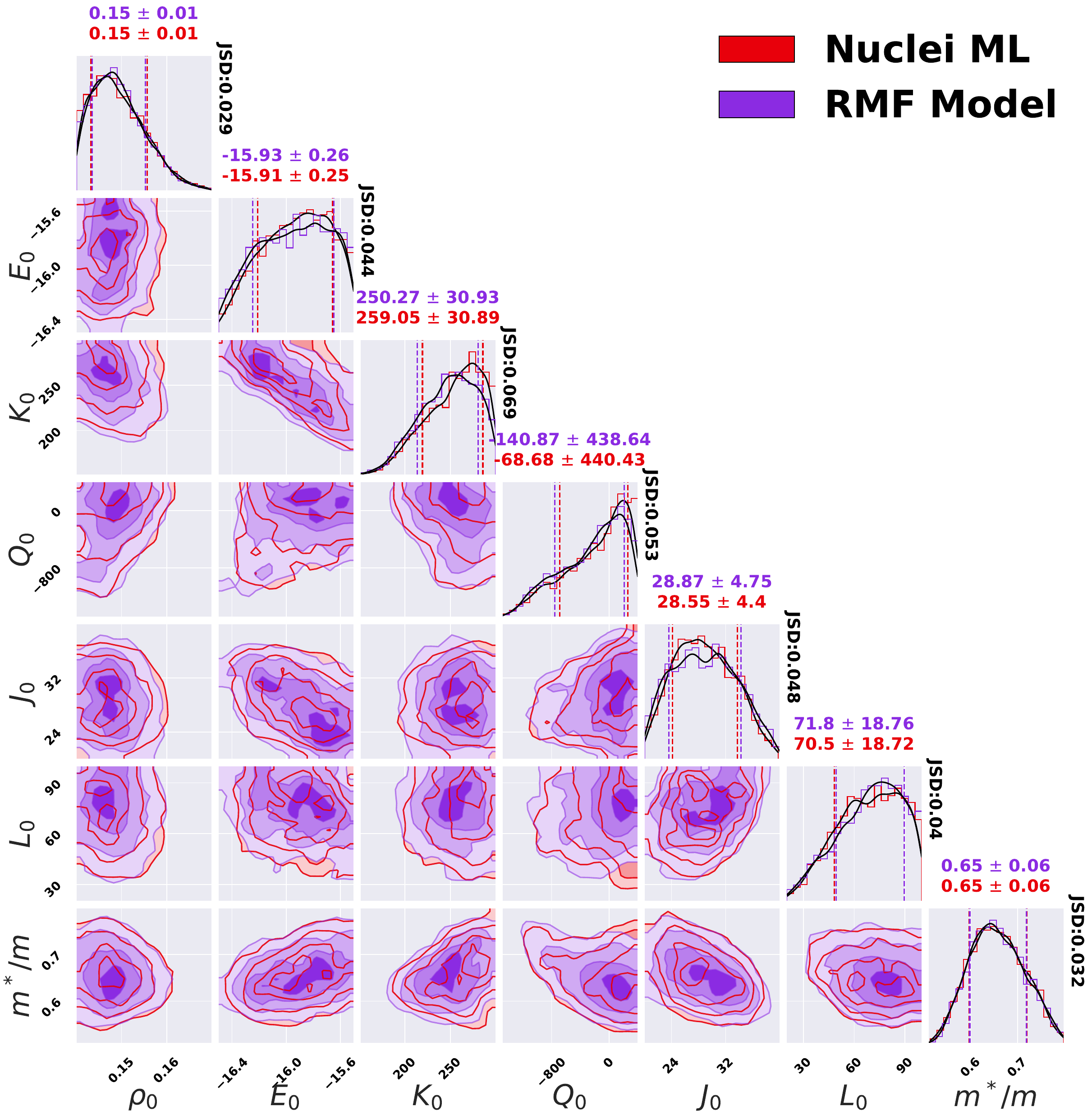}% Here is how to import EPS art
\caption{\label{fig:posterior}The posterior distribution of the Bayesian analyses using both the NML model and RMF model are presented. The median values for both the cases are shown above the diagonal plots, which also display the marginalized posterior distributions along with the 68\% CI marked by vertical dashed lines. The Jensen-Shannon divergence (JSD), which quantifies the difference between the parameter distributions for the two cases, is shown vertically along the marginalized posteriors.}
\end{figure}

The primary objective of this work is to employ NML as a computationally efficient alternative to RMF predictions of finite-nuclei properties, enabling their explicit inclusion in Bayesian analyses of neutron star equations of state. Previous studies \cite{Patra2023, venneti2024unraveling, Tsang2020} have shown that low-density constraints from finite nuclei strongly influence neutron-star properties and their correlations with NMPs. However, the high computational cost of mean-field calculations prevents such constraints from being systematically applied. In this study, we demonstrate the feasibility of NML in a  Bayesian analysis, highlighting its ability to reproduce its underlying microscopic model, RMF. This work is intended as a first step toward a broader program that incorporates both ground-state and excitation properties of nuclei in the context of nuclear astrophysics.

In the Bayesian analysis, we first sample the RMF parameters [see Eq. (\ref{lagrangian})] from the prior distributions specified in Table \ref{tab:rmfparameters}. Using the RMF formalism outlined in Sec. \ref{Sec1}, the posterior distribution of the NMPs is then generated by applying constraints on the binding energies and charge radii of $^{40}$Ca$_{20}$ and $^{208}$Pb$_{82}$ nuclei within the accuracy of 2\% of their experimental values. For this purpose we employ a Gaussian likelihood, defined as,
\begin{equation}
\log \mathcal{L}(\mu,\sigma_i^2 \mid \Theta)
= -\tfrac{1}{2}\sum_{i=1}^{n}\!\left[
\log\!\bigl(2\pi\sigma_i^{2}\bigr)
+ \frac{(\mu_i-y_i)^2}{\sigma_i^{2}}
\right],
\end{equation}
where $\mu$ are the experimental values of the FN observables, $y_i$ the corresponding RMF predictions, and $\sigma_i$ the associated uncertainties (taken as 2\% of $\mu_i$). We additionally generate another posterior distribution, this time employing the NML emulator to generate the corresponding predictions.

Fig.\ref{fig:posterior} presents the posterior distributions of the NMPs obtained from the Bayesian analysis for both the NML and RMF models. The diagonal plots display the median and 68\% CI of the marginal posterior distributions for the NMPs.
The two dimensional distributions, as well as the one dimensional marginalized posterior distributions given along the diagonal are very similar for the RMF and NML cases. We want to emphasize here that the similarity mentioned above ensures that the NML has successfully captured the underlying correlations of the RMF model while the model itself is being explored over a large range. To give a quantitative assessment, we calculate the JS divergence for these distributions of individual NMPs, as indicated alongside the diagonal plots in Fig.\ref{fig:posterior}. The good performance of NML is supported by these JS divergence values, all of which remain below 0.1.
The pairwise correlations among the NMPs are also faithfully reproduced by NML in the Bayesian sampling, in close agreement with those obtained from the RMF model. Such correlations, as highlighted in previous studies\cite{Patra2023, Malik2020,venneti2024unraveling}, play a central role in constraining nuclear matter properties. The ability of NML to capture not only the individual distributions of NMPs but also their mutual correlations is therefore a key strength of the approach. Further improvements can be achieved by training on a broader and more diverse set of nuclei. The computational cost is likewise expected to scale more favorably as additional nuclei are incorporated into the Bayesian analysis, a direction we reserve for future work with updated NML models. \modification{By employing the NML model in place of the RMF solver, the Bayesian sampling time is reduced by approximately three orders of magnitude ($\sim 10^3$), decreasing the runtime from about 4.5 hours to roughly 15 seconds. For a single evaluations, the computation of FN properties for a given parameter set exhibits an even larger acceleration, with a speedup factor of approximately $1.5 \times10^4$, reducing the evaluation time from about 2 seconds to 1.5 milliseconds.} The specifications of the computing system and neural network architecture used in this study are detailed in \ref{app_neural}. 

At this stage, it is worth reiterating the primary goal of the NML framework proposed here. When applied in a Bayesian analysis with explicit constraints from finite nuclei, the model is designed to explore wide parameter ranges while avoiding erroneous sampling of outliers. This flexibility also allows NML to be incorporated alongside complementary constraints from finite-nuclei studies \cite{Mondal:2022cva} or astrophysical observations. It should also be emphasized that our objective is not to construct an extremely precise global mass model, a direction in which machine learning has already been extensively applied in nuclear physics\cite{neufcourt2019neutron,murarka2022neutron, wu2021nuclear, gao2021machine, shelley2021new, niu2022nuclear, mumpower2022physically, lovell2022nuclear, Li:2023dsv1, dellen2024predicting, mumpower2023bayesian, wu2024machine, yuksel2024nuclear}. Rather, the purpose of NML is to enable large-scale Bayesian analyses incorporating explicit finite nuclear constraints into analyses of NS properties while remaining rooted in a microscopic description as well as retaining its computational efficiency.

\section{Summary and outlook}\label{Sec4}
We employed a machine learning framework to develop an  emulator of RMF theory achieving an improved computational performance, which enables incorporation of explicit finite nuclei constraints on neutron star equation of state \cite{venneti2024unraveling} in an efficient manner. A neural network model was trained using a large dataset generated from RMF theory calculations. The dataset includes  binding energies and charge radii for several nuclei obtained for a large set of nuclear matter parameters (NMPs).  The neural network architecture of NucleiML (NML) consists of a Classifier and a Regressor. The NML Classifier identifies whether the input parameters would yield convergent finite nuclei properties or not. The classifier achieved an accuracy of 95\% for categorizing inputs into two classes: admissible and non-admissible. The NML regressor predict the binding energies and charge radii for inputs classified as  \textit{admissible} by the classifier. Regressor achieved a deviation in the binding energies and charge radii less than $\sim 5\%$, for 95\% of the test dataset, indicating an accuracy reasonable for the purpose of incorporating the FN properties explicitly in a Bayesian analysis. The performance of NML regressor on unseen nuclei improves significantly when the training set includes more diverse nuclei, indicating a scope to extend the framework to larger sets of diverse nuclei.

The NML was incorporated into a Bayesian framework to constrain the NMPs based on FN experimental data. The results of the Bayesian analyses using NML matched closely those of the RMF model.  The Jensen-Shannon divergence values indicate high similarity between the marginalized posterior distributions for the NMPs considered from NML and the RMF model. NML also significantly accelerates computations compared to the RMF model, by nearly ten times less computational effort bringing the runtime down from hours to minutes.

The NML demonstrates how machine learning can efficiently replicate the predictive capabilities of computationally intensive RMF models. The substantial reduction in computational time makes NML a highly efficient alternative while maintaining accuracy, facilitating faster exploration of the parameter space in Bayesian inference. It provides a promising approach for accelerating nuclear property calculations and enables a more accessible integration of explicit FN and astrophysical constraints on the global behavior of EoS into Bayesian frameworks for uncertainty quantification and parameter optimization. Moreover, the improved performance of the regressor with the inclusion of more diverse nuclei in the training provides a strong motivation to extend this work in that direction. \modification{Furthermore, incorporating deformed nuclei will already a more stringent test of generalization at the mean-field level by enabling the emulator to learn features associated with collective deformation and long-range correlations beyond spherical systems. This framework can be also extended to other EDF families, such as Skyrme or Gogny by generating consistent training datasets and redefining the corresponding input parameter spaces while retaining the same surrogate architecture. We leave these ventures for future exploration.}

\ack{The authors acknowledge fruitful suggestions and discussions with Prasanta Char. BKA acknowledges with gratitude the support received under the Raja Ramanna Chair scheme of the Department of Atomic Energy (DAE), Government of India. AV acknowledges the CSIR-HRDG for support through the CSIR-JRF 09/1026(16303)/2023-EMR-I. CM acknowledges partial support
from the Fonds de la Recherche Scientifique (FNRS, Belgium) and the Research Foundation Flanders (FWO, Belgium) under the EOS Project nr O022818F and O000422.}

% \funding{Sample text inserted for demonstration.}
% This section is a list of funder names and grant numbers
% \sout{\modification{This framework can be extended to other EDF families by generating consistent training datasets and redefining the corresponding input parameter spaces while retaining the same surrogate architecture. In addition, incorporating deformed nuclei will provide a more stringent test of generalization by enabling the emulator to learn features associated with collective deformation and long-range correlations beyond spherical systems.}} 
\roles{
\begin{itemize}
    \item \textbf{Anagh Venneti}\orcid{0000-0002-0812-2702} 
Writing - original draft, investigation, formal analysis 
\item \textbf{Chiranjib Mondal}\orcid{0000-0002-9238-6144}
Writing - review and editing, investigation, formal analysis, conceptualization
\item \textbf{Sk Md Adil Imam}\orcid{0000-0003-3308-2615}
investigation 
\item \textbf{Sarmistha Banik}\orcid{0000-0003-0221-3651}
Writing - review and editing, supervision 
\item \textbf{Bijay K. Agrawal}\orcid{0000-0001-5032-9435}
Writing – review and editing, supervision, conceptualization
\end{itemize}}
% List author names and the contributions made to the article, using terms from the NISO Contributor Roles Taxonomy (CRediT) https://credit.niso.org

\data{The general architecture and other details of the neural network has been given in Figure.\ref{flowchart} and in \ref{app_neural}. The training dataset and the weights of \textit{NucleiML} will be made available on request.}
% For more information on IOP Publishing's research data policy see: https://publishingsupport.iopscience.iop.org/questions/research-data/

% \suppdata{Sample text inserted for demonstration.}

\appendix
\counterwithin{figure}{section}
\counterwithin{table}{section}
\renewcommand{\thesection}{Appendix \Alph{section}}
\renewcommand{\thefigure}{\Alph{section}\arabic{figure}}
\renewcommand{\thetable}{\Alph{section}\arabic{table}}

\setcounter{figure}{0}
\section{Neural Network Specifications and performance metrics}\label{app_neural}
We utilized the AMD EPYC 7452 32-core processor with an \verb|x86_64| architecture for training our neural networks. The processor operates at a base clock speed of 1.5 GHz with a boost clock of 2.35 GHz. This computational setup allowed us to efficiently train models on a dataset containing approximately 1,000,000 samples. For classification tasks, the NucleiML classifier employs the Rectified Linear Unit (ReLU) activation function in the inner layers, while the output layer uses a softmax activation with a normal kernel initializer. In regression tasks, the NML regressors also utilize ReLU activation in the inner layers, but the output layer employs a linear activation function. Both models are trained using a learning rate of $10^{-3}$ and Adam optimizer\cite{kingma2014adam}. Hyperparameter tuning is conducted via grid search, implemented using the Scikit-learn and \modification{SciKeras} Python packages \cite{scikit-learn}. The key configurations are summarized as follows:

\begin{table}[H]
\centering
\begin{tabular}{ll}
CPU              : & AMD EPYC 7452\\
Architecture     : & \verb|x86_64|\\
Cores            : & 32-cores, with 2 threads per core\\
Base Clock Speed : & 1.5 GHz \\
Max. Clock speed : & 2.35 GHz \\
Activation functions used : & ReLU (Classifier and Regressor)\\
& SoftMax (outer layer for Classifier)\\
& Linear (outer layer for Regressor)\\
\end{tabular}
\end{table}
In addition to the hyperparameters discussed in Sec. \ref{schematic}, several others play an important role. The batch size specifies the number of training samples used before updating the model weights. Larger batch sizes generally yield smoother training but may reduce generalization performance. For the NML framework, the optimal batch sizes were found to be 32 for the classifier and 128 for the regressor. Another class of hyperparameters governs the fitting behavior through regularization techniques, such as dropout rates or weight decay, which are commonly employed to prevent overfitting and improve generalization. However, the use of such regularization parameters in our case only led to underfitting, and we therefore did not employ them in the final analysis.

\modification{The dataset was partitioned into training, validation, and test sets in an 70–15–15 for the classifier (70-21-9 for the regressor) ratio using a fixed random seed to ensure reproducibility. Hyperparameter tuning was performed using three-fold cross-validation on the training set, with model selection based exclusively on validation performance. Training was conducted for a minimum of 600 epochs for classifier (300 for the regressor), with early stopping applied thereafter based on validation loss to prevent overfitting while ensuring adequate convergence. Throughout this procedure, careful separation of data subsets was maintained to prevent any form of data leakage, ensuring that neither validation nor test samples influenced model fitting or hyperparameter optimization.}

% \section*{References}
\section{\modification{The role of the classifier}}\label{class}
\subsection*{\modification{Conceptual role}}
\modification{The NML framework employs ther classifier as a admissibility filter, operating prior to the regressor. The primary motivation for this arises from the fact that not all sampled parameter sets represent physically meaningful or numerically converged RMF solutions. The regressor is trained to approximate the RMF predictions over admissible parameter space. However, this is not intrinsically enforced in its framework and this is where the classifier plays that role.} 

\modification{The classifier partitions the parameter space into admissible and non admissible regions. This separation ensures that the regressor only operates in the admissible regions where the mapping between the NMPs and FN observables is well defined and physically meaningful.}

\subsection*{\modification{Representative examples}}
\modification{Non admissibility of a given combinations of the NMPs arises from one of the following:}
\modification{\begin{itemize}
    \item \textbf{Non convergent :} RMF results that do not generate stable FN observables, even after 999 iterations.
    \item \textbf{No couplings generated :} Those combinations of NMPs that do not yield reasonable coupling parameters.
    \item \textbf{Threshold check :} FN observable lying beyond a threshold (20\% in our case) from experimental observation.
\end{itemize}}
\modification{Table.~\ref{tab:examples} present a representative set of examples with the respective reasons for their classifications.}
\begin{table}
\centering
\caption{Representative parameter sets classified as non-plausible, along with the corresponding reasons for their classification}
\label{tab:examples}
\begin{tabular}{lccc}
 & Set 1 & Set 2 & Set 3 \\
\hline
\hline
$r_0$              & 0.1482 & 0.1616 & 0.1698 \\
$E_0$              & -15.789 & -15.570 & -16.086 \\
$K_0$              & 205.889 & 299.820 & 226.976 \\
$Q_0$              & -1374.928 & -1394.497 & -644.027 \\
$m_{\mathrm{eff}}$ & 0.581 & 0.765 & 0.736 \\
$J_0$              & 20.326 & 38.108 & 39.218 \\
$L_0$              & 29.328 & 90.025 & 55.929 \\
$N$                & -- & 20 & 8 \\
$Z$                & -- & 20 & 8 \\
Flag               & Non-plausible & Non-plausible & Non-plausible \\
Reason             & No couplings generated & FN values not converged & FN values converged \\
BE                 & -- & -- & -101.349 \\
Iterations         & -- & 999 & 506 \\
\hline
\hline
\end{tabular}
\end{table}

\subsection*{\modification{Effect on the posterior distributions}}
\begin{figure}[h]
    \centering
    \includegraphics[width=0.95\linewidth]{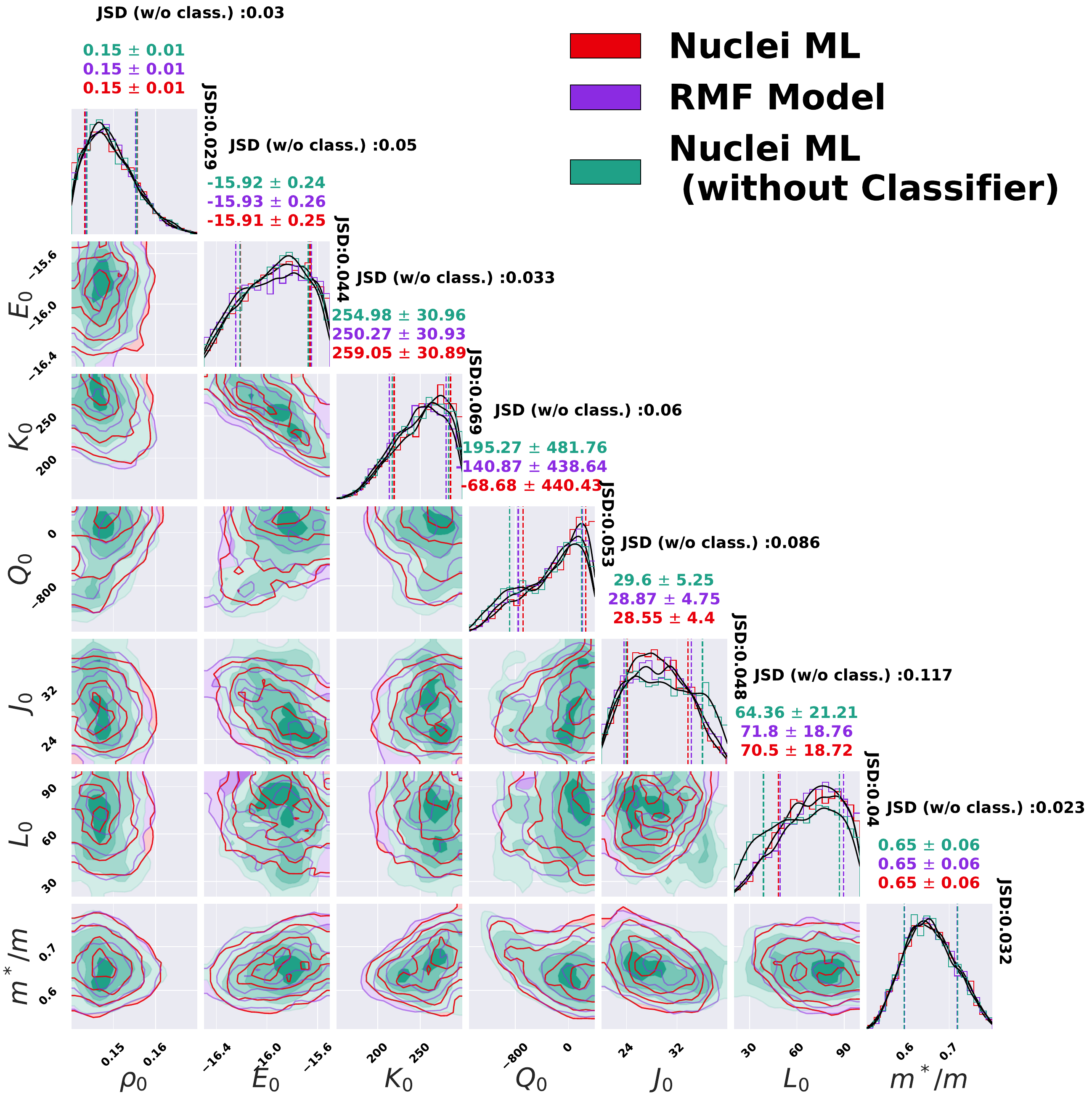}
    \caption{The Bayesian analysis with NucleiML (red), RMF model (violet), and NucleiML without the classifier (green).}
    \label{fig:1withoutclassifier}
\end{figure}

\modification{To quantitatively asses the impact of classifier on the setup of Bayesian inference, posteriors were computed with and without the admissibility filter. The posterior distributions and median values (shown in Fig.~\ref{fig:1withoutclassifier}) of $J_0$ and $L_0$  exhibit small but systematic shifts from those obtained with the RMF model. The corresponding JS divergence values quantify these differences and indicate a modest increase in divergence when the classifier is not employed. The credible intervals for certain parameters (notably $J_0$ and $L_0$) are moderately broader without the classifier. While the shifts observed without the classifier are moderate in magnitude and the credible intervals remain largely overlapping, these are systematic and measurably increase the JS divergence for certain parameters (notably $J_0$ and $L_0$). Since the aim of NucleiML is to serve as a faithful emulator of the underlying model, even such modest but reproducible deviations are undesirable. The inclusion of the classifier reduces these shifts and improves agreement with the posteriors obtained from the RMF model.}

\subsection*{\modification{Threshold Sensitivity analysis}}
\begin{figure}[htbp]
    \centering
    \begin{minipage}{0.5\textwidth}
        \centering
        \includegraphics[width=\textwidth]{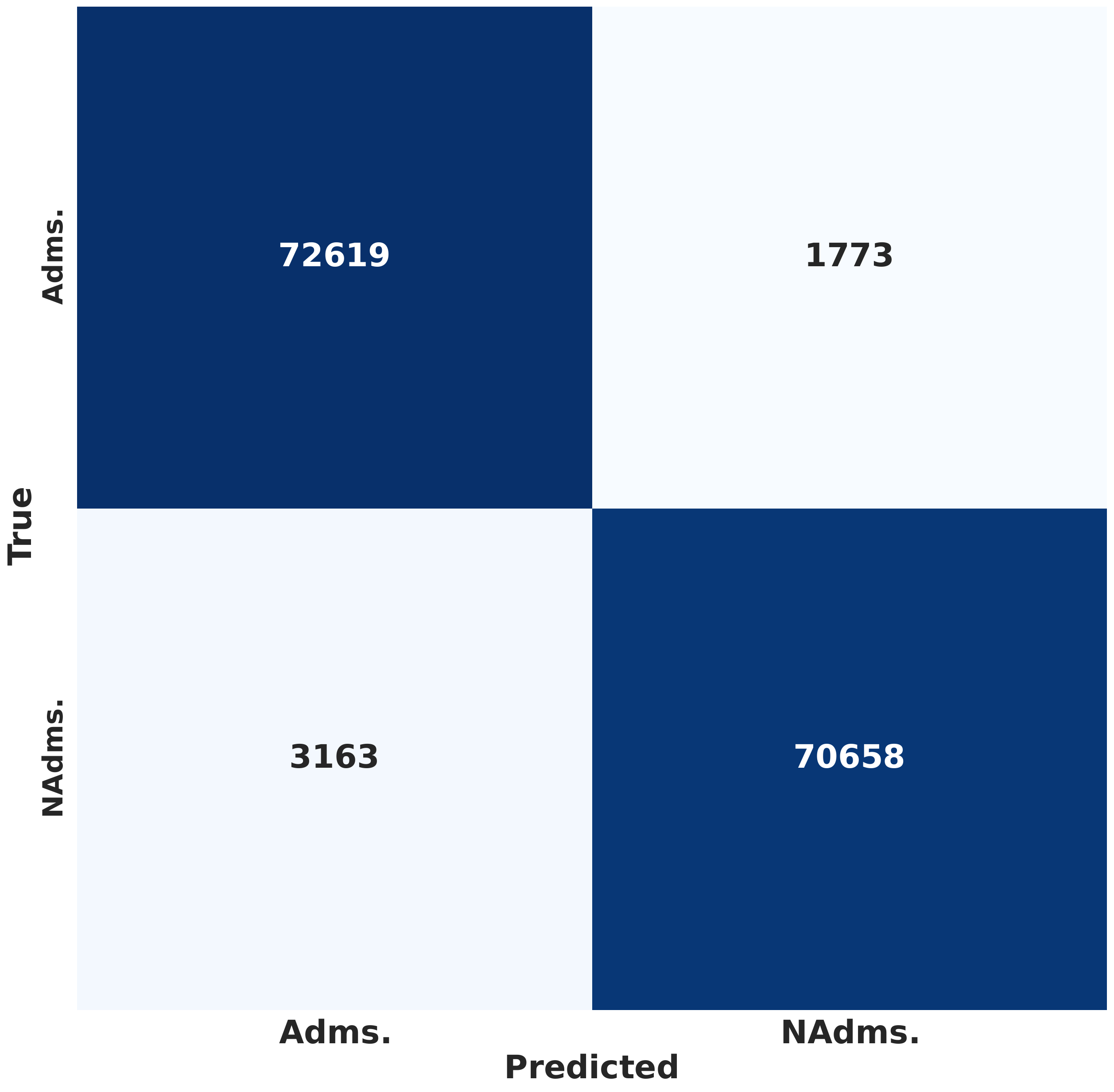}
        \caption{Threshold of 30\%}
    \end{minipage}
    \hfill
    \begin{minipage}{0.48\textwidth}
        \centering
        \includegraphics[width=\textwidth]{New_1_Confusion_Matrix_ver_2_diff.pdf}
        \caption{Threshold of 20\%, also reported in Fig.\ref{fig:confusion}}
    \end{minipage}
    
    \vspace{1.5cm}  % Adjust vertical spacing
    
    \begin{minipage}{0.5\textwidth}
        \centering
        \includegraphics[width=\textwidth]{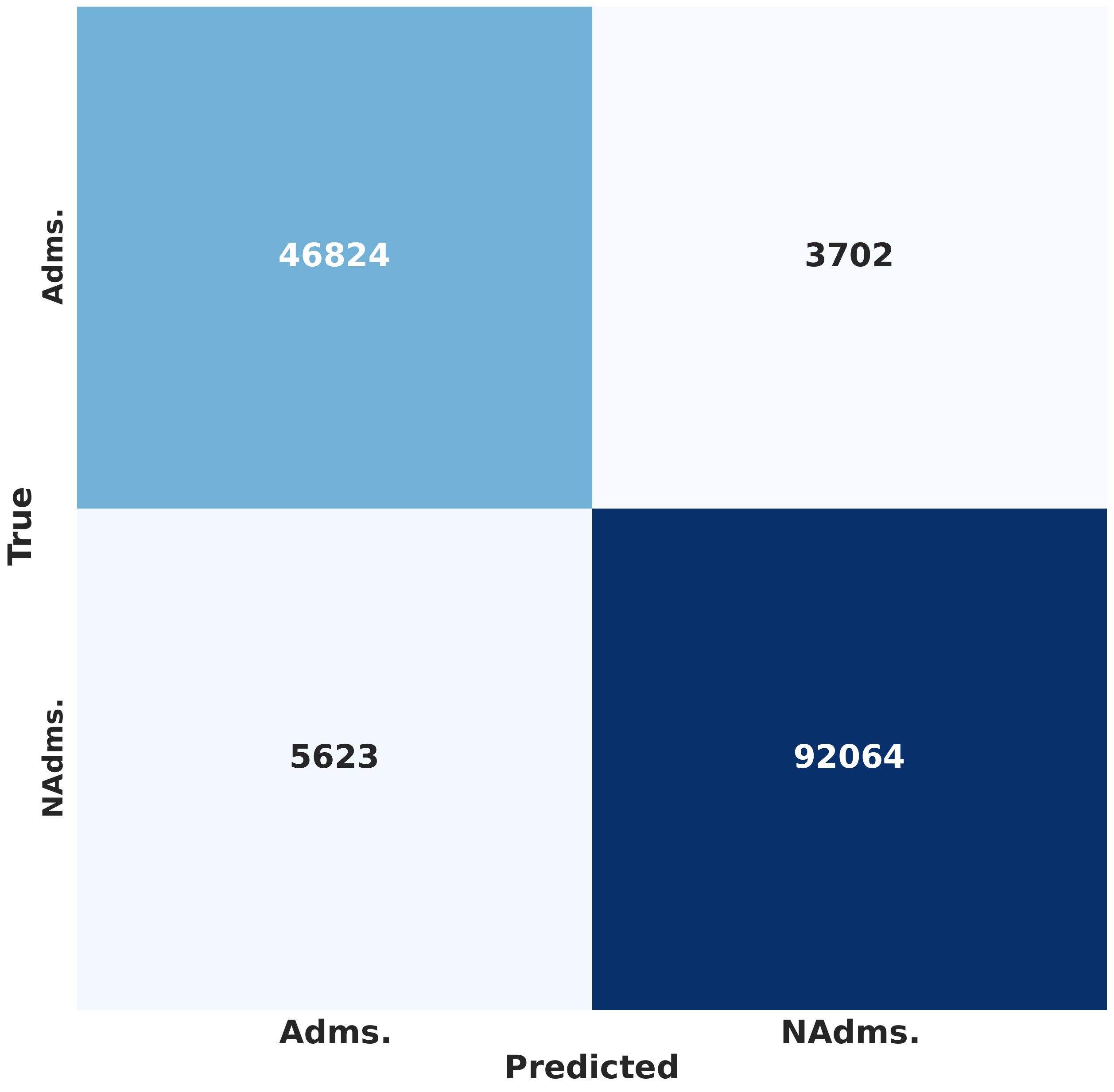}
        \caption{Threshold of 10\%}
    \end{minipage}
    \hfill
    \begin{minipage}{0.48\textwidth}
        \centering
        \includegraphics[width=\textwidth]{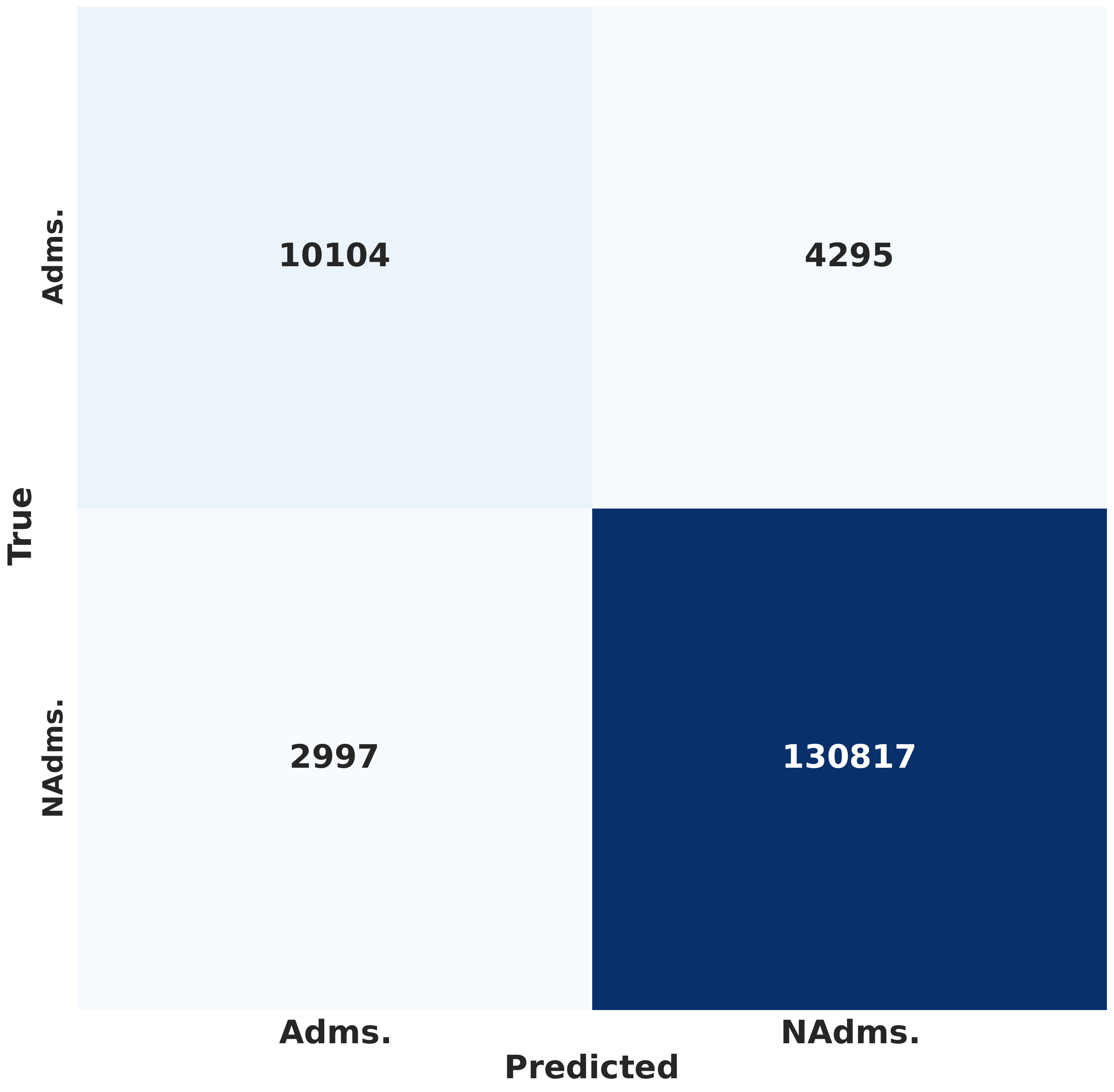}
        \caption{A fixed threshold of 10 MeV across the nuclei.}
    \end{minipage}

    \caption{\modification{The confusion matrices for the classifier performance with different thresholds.}}
    \label{fig:all_plots_class}
\end{figure}
\modification{The admissibility threshold, defined relative to experimental observables, determines the window within which a parameter set is admissible. Its selection represents a compromise between physical selectivity and numerical efficiency. A threshold that is overly restrictive may significantly narrow the admissible region, introduce class imbalance in the training dataset, and increase rejection rates during Bayesian sampling. Conversely, a permissive threshold may admit strongly unphysical parameter combinations, thereby degrading the physical interpretability of the posterior distributions. To assess the robustness of our chosen criterion (20\% relative threshold), we performed a systematic sensitivity analysis by varying the admissibility condition. Specifically, classifier performance was evaluated for relative thresholds of 30\%, 20\% (reported in the main manuscript), 10\%, and for a fixed absolute threshold of 10 MeV. The resulting accuracy, precision, and recall for both plausible and non-plausible classes are summarized in Table~\ref{tab:threshold_sensitivity_transposed}. The confusion matrices are shown in Fig.~\ref{fig:all_plots_class}.}
\begin{figure}[h]
\centering
\includegraphics[width=\textwidth]{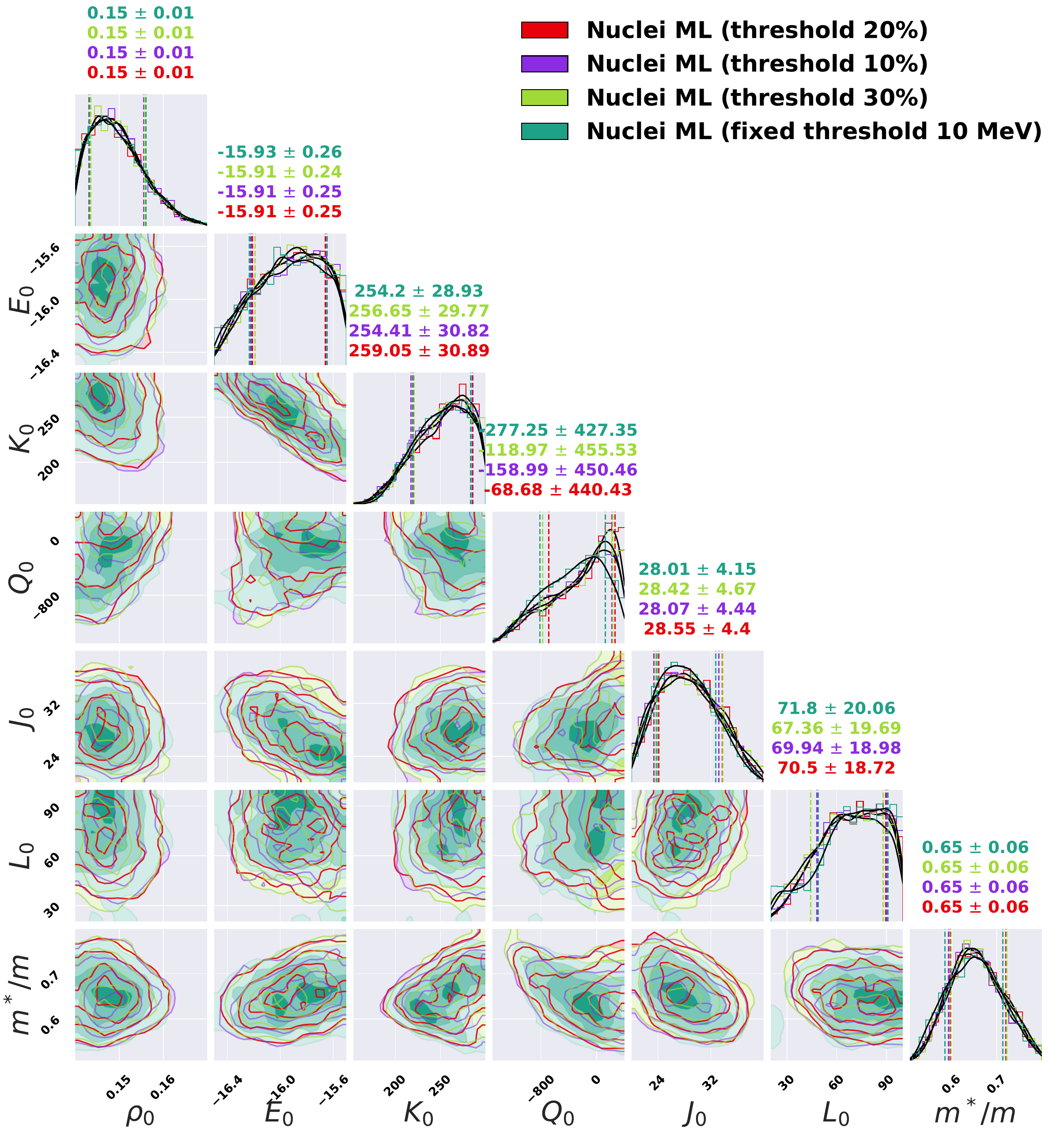}% Here is how to import EPS art
\caption{\label{fig:posterior}The posterior distribution of the Bayesian analyses using different thresholds are presented. The median values for both the cases are shown above the diagonal plots, which also display the marginalized posterior distributions along with the 68\% CI marked by vertical dashed lines.}
\end{figure}

\begin{table}[h]
\centering
\caption{Sensitivity of classifier performance to the admissibility threshold.}
\label{tab:threshold_sensitivity_transposed}
\begin{tabular}{lcccc}
\hline
\hline
Metric 
& 30\% (relative) 
& 20\% (relative) 
& 10\% (relative) 
& 10 MeV (absolute) \\
\hline
Accuracy 
& 0.9667 
& 0.9521 
& 0.9371 
& 0.9508 \\

Precision (Plausible) 
& 0.9583 
& 0.9341 
& 0.8928 
& 0.7712 \\

Recall (Plausible) 
& 0.9762 
& 0.9661 
& 0.9267 
& 0.7017 \\

Precision (Non-Plausible) 
& 0.9755 
& 0.9691 
& 0.9613 
& 0.9682 \\

Recall (Non-Plausible) 
& 0.9572 
& 0.9398 
& 0.9424 
& 0.9776 \\
\hline
\hline
\end{tabular}
\end{table}

\modification{The overall classification accuracy remains high across all threshold choices, exceeding 93\% in each case. For relative thresholds (30\%, 20\%, 10\%), precision and recall decrease gradually as the criterion becomes more restrictive, reflecting the increasing difficulty of separating admissible and non-admissible regions under tighter physical constraints. Importantly, the 20\% and 30\% thresholds maintain a relatively balanced trade-off between precision and recall for both classes. In contrast, the fixed 10 MeV absolute threshold introduces a pronounced asymmetry in precision and recall. This behavior is primarily attributed to strong class imbalance induced by the absolute admissibility criterion, which substantially reduces the number of plausible samples. As a consequence, the classifier exhibits degraded recall for the plausible class while maintaining high recall for the non-plausible class.}

\modification{Beyond classification metrics, the impact of threshold variation on Bayesian inference was also examined (see Fig. \ref{fig:posterior}). The posterior distributions corresponding to each threshold case remain statistically consistent, with no evidence of systematic bias or artificial concentration in parameter space. While mild quantitative differences in spread are observed, the global structure of the inferred NMP distributions remains stable. This indicates that the inference mechanism is robust against reasonable variations of the admissibility threshold.}

\modification{Computational efficiency was also assessed. For relative thresholds (30\%, 20\%, 10\%), sampling runtimes remain comparable (approximately 15–16 seconds). However, the absolute 10 MeV threshold nearly doubles the runtime (approximately 30 seconds), reflecting the significantly narrowed admissible region and the corresponding increase in sampling rejection rates. Overall, this analysis demonstrates that relative thresholding preserves both physical interpretability and computational efficiency. Among the cases considered, the 20\% relative criterion provides a conservative yet stable choice, yielding balanced classifier performance, robust posterior inference, and efficient Bayesian sampling.}

\section{\modification{Performance based on individual nucleus}}\label{nuc_wise_rel_err}
\subsection*{Relative error for trained nuclei}
\begin{figure}[htbp]
    \centering
    \begin{minipage}{0.48\textwidth}
        \centering
        \includegraphics[width=\textwidth]{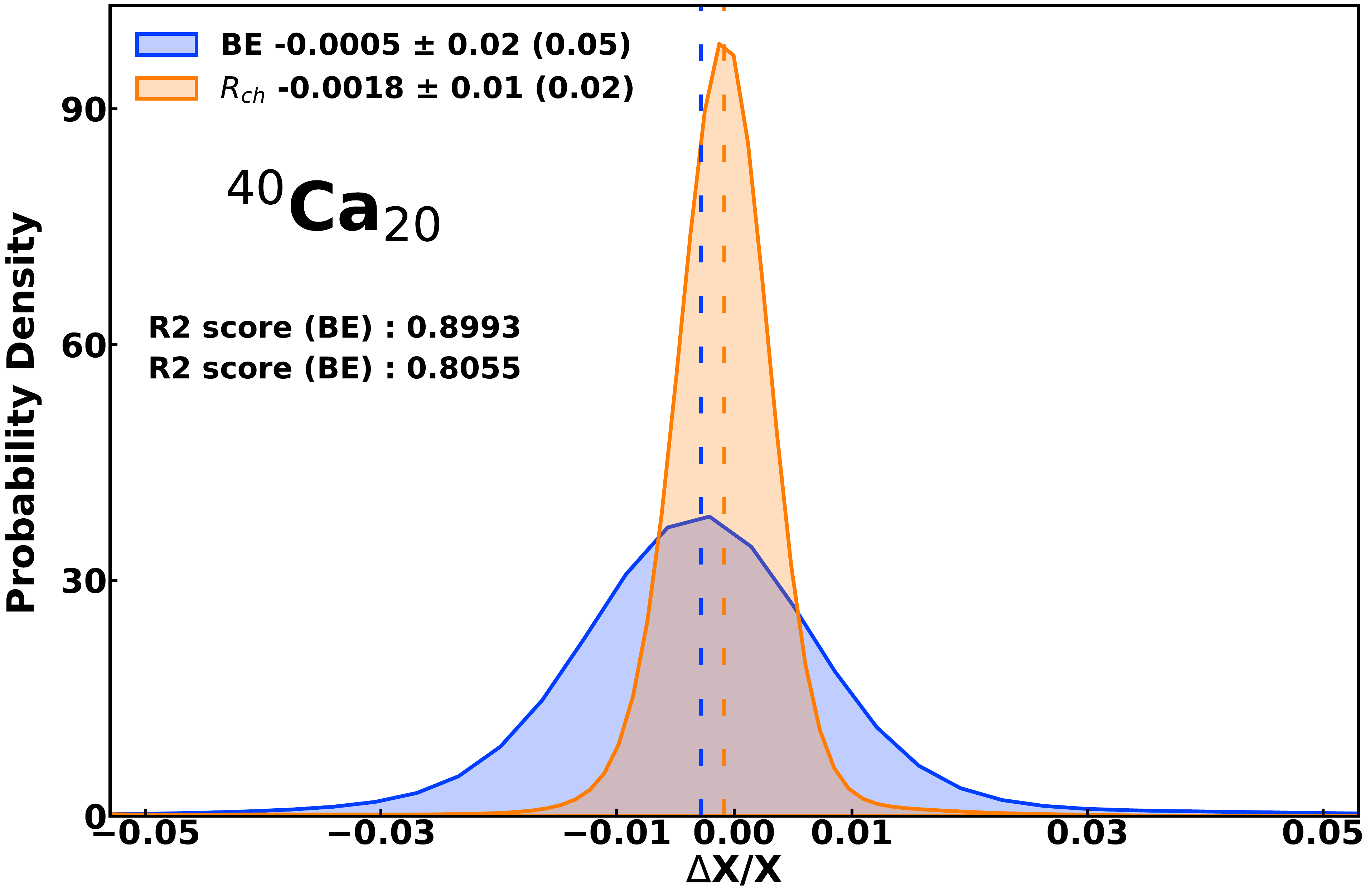}
        % \caption{Threshold of 30\%}
    \end{minipage}
    \hfill
    \begin{minipage}{0.48\textwidth}
        \centering
        \includegraphics[width=\textwidth]{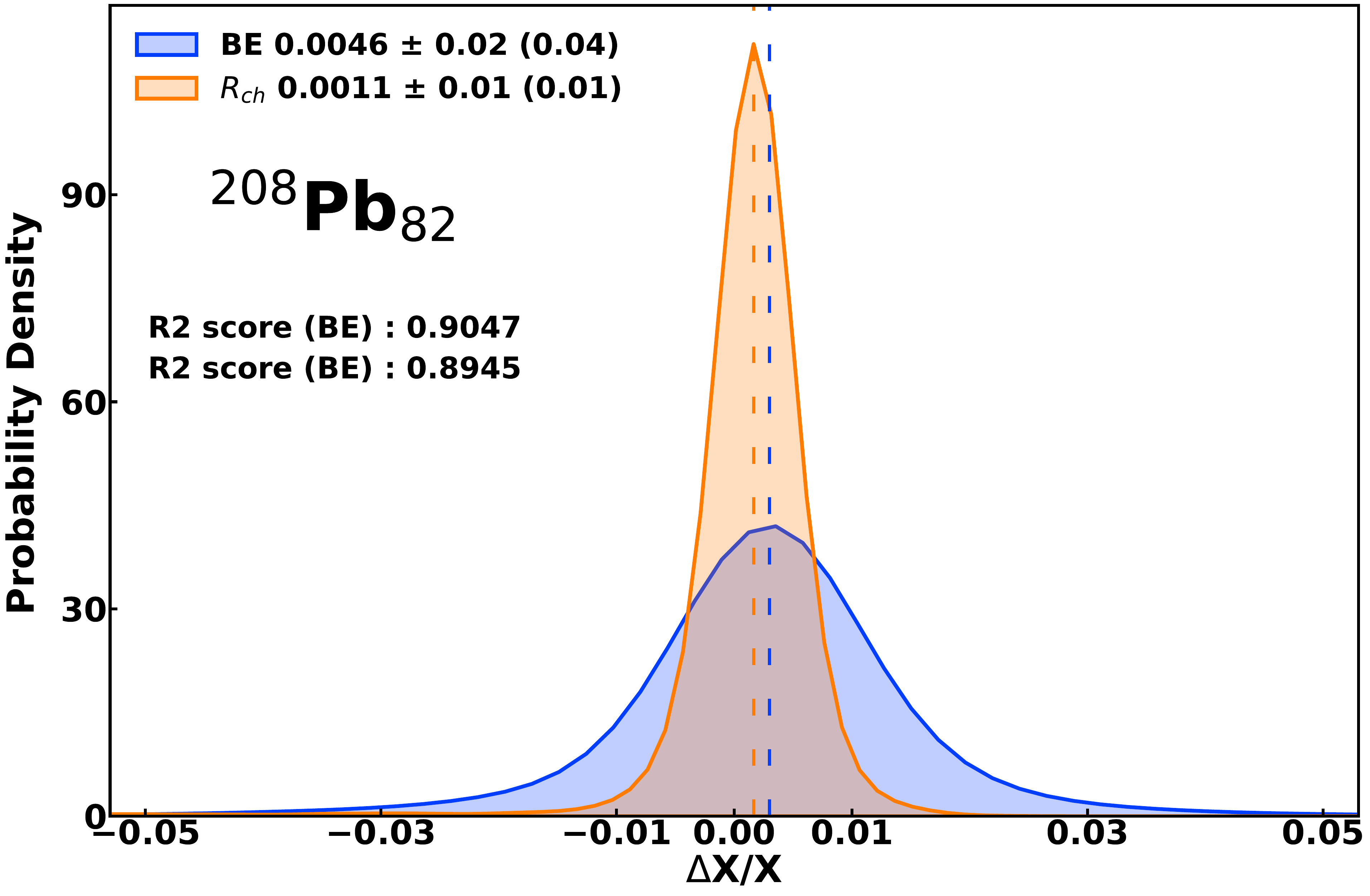}
        % \caption{Threshold of 20\%, also reported in Fig.\ref{fig:confusion}}
    \end{minipage}
    \caption{\modification{Relative errors ($\Delta X/X$, Eq.\ref{dev}) for for binding energy and charge radius of $^{40}$Ca$_{20}$ and $^{208}$Pb$_{82}$ }}
    \label{fig:rel_error}
\end{figure}
\modification{The figure shows the probability density distributions of the relative errors, $\Delta X / X$, for the binding energy (BE) and charge radius ($R_{\mathrm{ch}}$) of $^{40}\mathrm{Ca}_{20}$ and $^{208}\mathrm{Pb}_{82}$. In both panels, the distributions are sharply peaked around zero, indicating that the NML surrogate produces essentially unbiased predictions. The dashed vertical lines denote the mean relative errors, while the legend reports the mean and standard deviation (with the value in parentheses indicating the 95 \% CI). For both nuclei, the charge radius errors exhibit a noticeably narrower spread compared to the binding energy, reflecting reduced variance and greater stability in radius predictions. The heavier nucleus $^{208}\mathrm{Pb}_{82}$ shows slightly tighter distributions overall, suggesting improved robustness in the heavy-mass region. The consistently high $R^2$ scores shown in each panel further confirm the strong agreement between the surrogate predictions and the reference RMF calculations for these benchmark systems.}
\subsection*{R2 scores for unseen nuclei}
\modification{Table~\ref{tab:r2_unseen_nuclei} summarizes the extrapolation performance of the NML surrogate for unseen nuclei in terms of the coefficient of determination ($R^2$) for both binding energy and charge radius. The Initial Set exhibits limited predictive capability for several nuclei, with strongly negative $R^2$ values in some cases (e.g., $^{56}\mathrm{Ni}{28}$ and $^{100}\mathrm{Sn}{50}$ for binding energy), indicating poor generalization outside the training domain. The progressive augmentation of the training data (Initial + I and Initial + II) leads to a marked and systematic improvement in extrapolation accuracy, particularly for binding energies, where most nuclei achieve positive and often high $R^2$ values under the Initial + II configuration. For charge radii, improvements are also observed, although the behavior is more nucleus-dependent and reflects the greater sensitivity of radius predictions to data coverage. Overall, the results demonstrate that targeted dataset enrichment substantially enhances the surrogate’s robustness and stabilizes predictions in regions of parameter space not explicitly sampled during training.}
\begin{table*}[h]
\centering
\caption{R$^2$ scores for unseen nuclei for binding energy and charge radius across different training configurations. Negative values indicate poor extrapolation performance.}
\label{tab:r2_unseen_nuclei}
\begin{tabular}{l|ccc|ccc}
\hline
\hline
 & \multicolumn{3}{c|}{Binding Energy ($R^2$)} 
 & \multicolumn{3}{c}{Charge Radius ($R^2$)} \\
Nucleus 
 & Initial Set 
 & Initial + I 
 & Initial + II 
 & Initial Set 
 & Initial + I 
 & Initial + II \\
\hline

$^{30}\mathrm{Ne}_{10}$ 
 & 0.1342 & 0.7662 & 0.6297 
 & -0.2837 & -0.6734 & -3.0926 \\

$^{54}\mathrm{Ca}_{20}$ 
 & 0.1871 & 0.8840 & 0.9005 
 & 0.6199 & 0.7072 & 0.6906 \\

$^{56}\mathrm{Ni}_{28}$ 
 & -1.8191 & -0.3130 & 0.0242 
 & 0.7626 & 0.6407 & -0.9842 \\

$^{100}\mathrm{Sn}_{50}$ 
 & -0.5571 & -1.9090 & 0.0697 
 & -0.3953 & 0.5338 & 0.3313 \\

$^{116}\mathrm{Sn}_{50}$ 
 & 0.8300 & 0.5185 & 0.9019 
 & 0.4634 & 0.7097 & 0.9023 \\

$^{138}\mathrm{Sn}_{50}$ 
 & 0.8410 & 0.8589 & 0.8442 
 & 0.7246 & 0.7155 & 0.6805 \\

$^{144}\mathrm{Sm}_{62}$ 
 & 0.7209 & 0.5268 & 0.8239 
 & 0.3301 & 0.5170 & 0.5881 \\

\hline
\hline
\end{tabular}
\end{table*}
\newpage
\bibliographystyle{iopart-num}   % if provided in the package
\bibliography{References}
\end{document}